%% file: paper.tex
\documentclass[hyper,11pt,paper]{article}
\usepackage{jheppub}
\usepackage[usenames,dvipsnames,table]{xcolor}
\usepackage{graphicx,amsmath,amssymb,multirow,array,bm,mathrsfs,bbold,bbm}
\usepackage{appendix}
\usepackage{footmisc}
\usepackage{epsf,amsfonts,slashed,soul}
\usepackage{lipsum}
\usepackage{subfig}
\usepackage{pifont}
\usepackage{bbold}
\usepackage{dsfont}
\usepackage{tikz}
\usepackage{verbatim}
\usepackage{placeins}
\allowdisplaybreaks
 
\newcommand{\bea}{\begin{eqnarray}}
\newcommand{\eea}{\end{eqnarray}}

\input{macros}

\title{Virasoro Entanglement Berry Phases}

\author[a]{Jan de Boer,}
\author[a]{Ricardo Esp\'indola,}
\author[a]{Bahman Najian,}
\author[b]{Dimitrios Patramanis,}
\author[a]{\\Jeremy van der Heijden,}
\author[a]{and Claire Zukowski}
\affiliation[a]{Institute for Theoretical Physics, University of Amsterdam, Science Park 904, Postbus 94485, 1090 GL Amsterdam, The Netherlands}

\affiliation[b]{Faculty of Physics, University of Warsaw, ul. Pasteura 5, 02-093 Warsaw, Poland}

\emailAdd{J.deBoer@uva.nl}
\emailAdd{r.espindolaromero@uva.nl}
\emailAdd{b.najian@uva.nl}
\emailAdd{d.patramanis@uw.edu.pl}
\emailAdd{j.j.vanderheijden2@uva.nl}
\emailAdd{c.e.zukowski@uva.nl}

\abstract{We study the parallel transport of modular Hamiltonians encoding entanglement properties of a state. In the case of 2d CFT, we consider a change of state through action with a suitable diffeomorphism on the circle: one that  diagonalizes the adjoint action of the modular Hamiltonian. These vector fields exhibit kinks at the interval boundary, thus together with their central extension they differ from usual elements of the Virasoro algebra. The Berry curvature associated to state-changing parallel transport is the Kirillov-Kostant symplectic form on an associated coadjoint orbit, one which differs appreciably from known Virasoro orbits. We find that the boundary parallel transport process computes a bulk symplectic form for a Euclidean geometry obtained from the backreaction of a cosmic brane, with Dirichlet boundary conditions at the location of the brane. We propose that this gives a reasonable definition for the symplectic form on an entanglement wedge.}

\begin{document}
\maketitle

\input{introduction}
\input{berry}
\input{Section3bis}
\input{coadjoint}
\input{gravity}

\input{discussion}
\section*{Acknowledgements}
It is a pleasure to thank Ben Freivogel for initial collaboration. We also thank Raf Bocklandt, Bowen Chen, Bartlomiej Czech, Jackson R. Fliss, Victor Godet, Kurt Hinterbichler, Jani Kastikainen, Esko Keski-Vakkuri, Lampros Lamprou, Sergey Shadrin, Erik Verlinde and Zi-zhi Wang for discussions. DP is supported by NAWA “Polish Returns 2019.” RE and CZ are supported by the ERC Consolidator Grant QUANTIVIOL. BN is supported by the Spinoza Grant of the Dutch Science Organisation (NWO). JdB and JvdH are supported by the European Research Council under the European Unions Seventh Framework Programme (FP7/2007-2013), ERC Grant agreement ADG 834878. This work is supported by the
Delta ITP consortium, a program of the Netherlands Organisation for Scientific Research (NWO) that is funded
by the Dutch Ministry of Education, Culture and Science (OCW).

\appendix
\input{appendix}

\bibliographystyle{JHEP}
\bibliography{virasoro.bib}

\end{document}

%% file: Macros.tex
\newcommand{\be}{\begin{equation}}
\newcommand{\ee}{\end{equation}}
\newcommand{\ba}{\begin{eqnarray}}
\newcommand{\ea}{\end{eqnarray}}

\def\ket#1{| #1 \rangle}
\def\bra#1{\langle  #1 |}


%% file: introduction.tex
\section{Introduction}

A particular goal of holography is to understand the emergence of geometry from the boundary conformal field theory. Recent applications of quantum information theory in holography have given a means of directly probing geometry of the bulk, and thus have provided a promising avenue for addressing this question. 

One geometrical application of entanglement is an auxiliary space for holography known as kinematic space, which can be defined as the space of pairs of spacelike points in a CFT$_d$~\cite{Czech:2016xec, deBoer:2016pqk}. Perturbations of entanglement entropy are seen to propagate as fields on this space~\cite{deBoer:2015kda}. For CFT$_2$, kinematic space can additionally be obtained from the set of entanglement entropies associated to intervals~\cite{Czech:2015qta}. While fixed by the asymptotic conformal symmetry, kinematic space provides a tool for the reconstruction of bulk geometry in certain sufficiently symmetrical and controlled settings. For instance, it reconstructs geometry for locally AdS$_3$ spacetimes~\cite{Asplund:2016koz}. It also probes the geometry only outside of entanglement shadow regions that are inaccessible to spacelike geodesics~\cite{Freivogel:2014lja, Espindola:2018ozt}. This auxiliary space is a symplectic manifold, specifically it is a particular coadjoint orbit of the conformal group~\cite{Penna:2018xqq}.

The drawback here is of course the reliance on symmetries and special geometries. Is it possible to use entanglement to probe more general geometries? To this end, transport for $2$d kinematic space was generalized to a parallel transport process for the modular Hamiltonian~\cite{Czech:2017zfq, Czech:2019vih}.\footnote{For approaches to general reconstruction using null surfaces rather than spacelike extremal surfaces, see~\cite{Maldacena:2015iua, Engelhardt:2016wgb, Engelhardt:2016crc}. To move beyond entanglement shadow regions and geodesic barriers~\cite{Engelhardt:2013tra} one could also use timelike geodesics as probes. These are dual to circuit complexity as defined by the Nielsen geometric approach. They describe an auxiliary symplectic geometry which is also a coadjoint orbit of the conformal group, just a different one than kinematic space~\cite{Chagnet:2021uvi}.} In this setup, there is an associated Berry connection on kinematic space that computes lengths of curves in the bulk. More generally, a modular Berry connection can be shown to relate frames for CFT algebras associated to different states and subregions. Entanglement provides a connection that sews together nearby entanglement wedges and probes the geometry near the extremal surface. This connection builds spacetime from entanglement, reminiscent of the ER=EPR proposal~\cite{Maldacena:2013xja}. While the modular Hamiltonian admits a particularly simple, local description only in special cases, the parallel transport of modular Hamiltonians is true more generally, and its bulk description relies only on leading order in $1/N$ and sufficient smoothness of the extremal surface.

The parallel transport of modular Hamiltonians has been studied in the setting where the interval shape is varied, which connects to kinematic space~\cite{Czech:2016xec}. Shape-changing parallel transport has also been applied to study cases in holography where the modular chaos bound is saturated, which is governed by a certain algebra of modular scrambling modes that generate null deformations close to the extremal surface~\cite{DeBoer:2019kdj}. We are interested in generalizing beyond the case where the shape or interval location is varied, to consider modular parallel transport governed by a change of global \emph{state} (see also~\cite{Kirklin:2019ror} for a similar approach). For instance, one could imagine acting on a CFT on the cylinder by a large diffeomorphism contained in the Virasoro algebra. This would modify the algebra of operators on the interval. The redundancy by certain symmetries known as \emph{modular zero modes} which change the algebra but leave physical observables fixed results in a connection and non-trivial parallel transport, even in the case where the interval remains fixed. A general modular transport problem would consist of an amalgamation of these two kinds of parallel transport, with a simultaneous modification of both the state and interval shape.

Ultimately, we consider special transformations which do not lie in the Virasoro algebra as typically defined since they are not analytic, rather they vanish at the interval endpoints and are non-differentiable at these points. The reason for this is technical: to uniquely isolate the zero mode contribution it is necessary to have a decomposition into kernel and image of the adjoint action of the modular Hamiltonian. As we explain in Appendix~\ref{sec:NonDiagonalization}, this is not possible for the Virasoro algebra. This is a subtlety that, to our knowledge, has not been previously studied. For a large class of transformations which obey certain properties, we derive a general expression for the Berry curvature in Appendix~\ref{app:general}. We also explain how these non-standard vector fields have a simple interpretation as plane waves in the hyperbolic black hole geometry using the map of Casini, Huerta and Myers~\cite{Casini:2011kv}.

We define a suitable algebra of vector fields on the circle constructed from wave packets of these eigenstates. Much as similar group-theoretic parallel transport problems are governed by the geometry of symplectic manifolds known as coadjoint orbits, here that is the case as well. We show that the Berry curvature for state-changing parallel transport is equal to the Kirillov-Kostant symplectic form on an orbit associated to this algebra of vector fields.

State-changing parallel transport can also be related to bulk geometry. This has the advantage of accessing different geometrical data in the bulk, compared to the setting where only the interval shape is varied. We find that the Berry curvature for a fixed interval and changing state computes the symplectic form for a Euclidean conical singularity geometry obtained from the backreaction of a cosmic brane, subject to a suitable principal value prescription for regulating divergences near the interval endpoint. To match the curvature, we must impose Dirichlet boundary conditions at the location of the extremal surface. We interpret this as describing (and defining) a symplectic form associated to the entanglement wedge. In the discussion, we connect to earlier work on the holographic interpretation of the Berry curvature, and comment on the relation to the entanglement wedge symplectic form in the case of operator-based parallel transport. 

Modular parallel transport, either in the case of a changing shape or a changing state, is a parallel transport of \emph{operators} and density matrices. It is distinct from existing algebraic applications of parallel transport of \emph{states}, which for instance transform under unitary representations of a symmetry group. As part of this work we hope to clarify some of the differences, as well as various applications of each. In particular, we both review how kinematic space for CFT$_2$ can be understood in the language of operator-based parallel transport in Section~\ref{sec:states}, while also providing a new derivation of this same kinematic space using state-based parallel transport in Appendix~\ref{app:kinematic}. This gives two different ways of viewing the same problem, both utilizing group theory, reminiscent of the `Heisenberg' versus `Schr\"{o}dinger' pictures for quantum mechanics.\\

\noindent {\bf Outline}: We begin in Section~\ref{sec:berry} by giving a summary of both state and operator-based parallel transport, and providing a few examples of each. In Section~\ref{sec:transport}, we derive the boundary parallel transport process for transformations that diagonalize the adjoint action and compute the curvature in an example. We go into further detail in Section~\ref{sec:coadjoint} about the algebraic structure and the connection to coadjoint orbits. In Section~\ref{sec:bulk}, we present our proposal for the bulk dual using the symplectic form for Euclidean conical singularity solutions created from the backreaction of a cosmic brane. We end with a discussion about some subtleties and suggest future research directions. In Appendix~\ref{app:kinematic}, we provide a derivation of kinematic space using state-based parallel transport, and in Appendix~\ref{app:general} we derive a general expression for the curvature for operator-based parallel transport, which applies for any algebra. Finally, in Appendix~\ref{sec:NonDiagonalization} we discuss some subtleties about diagonalization of the adjoint action for the Virasoro algebra.

%% file: berry.tex
\section{Geometric Berry phases}\label{sec:berry}

Geometric phases can arise in quantum mechanics when a Hamiltonian depends continuously on certain parameters, such as an external magnetic field. This results in a state that differs from the starting state by a phase under a closed path in parameter space. Several generalizations of this notion have recently arisen in studies of conformal field theory and holography, relying for instance on the fact that entanglement can act as a connection that relates the Hilbert spaces of different subsystems. 

The applications to holography utilize group-based generalizations of the familiar geometric phases of quantum mechanics. In this section, we will review the tools that are relevant, making a distinction between two different approaches for group-based parallel transport depending on whether it is applied to states (a Schr\"{o}dinger-type picture) or density matrices (a Heisenberg approach). Before moving on to new results, we give some examples of how these different approaches have so far been applied to holography.

\subsection{States}\label{sec:states}
We begin by describing the parallel transport of states that transform under a unitary representation of a group (see~\cite{Oblak:2017ect} for applications to the Virasoro group). The basic idea is to generalize beyond a path in a space of parameters, as in quantum mechanics, to a path in a group representation. A gauge connection can be defined relating different tangent spaces along the path. If some unitaries in the representation act trivially on a starting state, this constitutes a redundancy by which the state may not return to itself under a closed path through the group manifold.

Specifically, consider a group $G$ with Lie algebra $\mathfrak{g}$, and a unitary representation $\mathcal{D}$ which acts on a Hilbert space $\mathcal{H}$. Take a state $\ket{\phi}\in\mathcal{H}$ that is an eigenstate of all elements in a  `stabilizer' subalgebra $\mathfrak{h}\subset \mathfrak{g}$, or equivalently it is left invariant up to a phase under the action of the corresponding subgroup $H\subset G$. Let $U(\gamma(t))\in \mathcal{D}$ with $\gamma(t)\in G$, $t\in [0,T]$ be a continuous path through this representation, which corresponds to a continuous path of states $\ket{\phi(t)} = U(\gamma(t))\ket{\phi}$. The states $\ket{\phi(t)}$ for all $\gamma(t)$ are often called generalized coherent states, and they parametrize the coset space $G/H$~\cite{Perelomov:1986tf,Yaffe:1981vf}.

The Berry connection is defined as
\be A = i \bra{\phi(t)} d \ket{\phi(t)} = i \bra{\phi} U^{-1} d U \ket{\phi}~,\ee
where $d$ is the exterior derivative on the group manifold, and we have used $U^\dagger = U^{-1}$ since the representation is unitary. The connection is just $A = i \bra{\phi} \mathcal{D}(\Theta) \ket{\phi}$ with $\Theta$ the Maurer-Cartan form associated to the group, $\Theta(\dot{\gamma}(t)) = \left.\frac{d}{d\tau}\right|_{\tau=t}[\gamma(t)^{-1} \gamma(\tau)]$. Under action by an element of the stabilizer subgroup, the state changes by a phase $\ket{\phi(t)} \rightarrow e^{i \alpha} \ket{\phi(t)}$. The connection then transforms as a gauge field, $A \rightarrow A -d\alpha$.

The associated Berry curvature is
\be F = dA~, \ee
and the geometric phase is defined as
\be \theta(\gamma) = \int_\gamma A~.\ee
This phase is in general gauge dependent, but is gauge invariant when the path $\gamma$ is closed. In this case, we can write

\be \theta(\gamma) = \oint_\gamma A~ = \int_{B | \partial B = \gamma} F~,\ee
where in the last line we have used Stokes' theorem to convert this to the flux of the Berry curvature over any surface $B$ with boundary $\gamma$. This measures the phase picked up by the state $\ket{\phi}$ under a closed trajectory through the group representation.

Similar techniques are relevant in the study of Nielsen complexity, which describes the geometry of the space of states related by unitaries, starting from a given reference state. A specific path through unitaries is known as a `circuit.' In conformal field theory, one can choose a reference state such as a primary that is invariant under a subset of the conformal symmetry. Defining the complexity further requires a notion of distance between states. Certain choices have relations to the Berry connection or curvature of state-based parallel transport~\cite{Caputa:2018kdj,Bueno:2019ajd,Akal:2019hxa,Erdmenger:2020sup,Flory:2020eot,Flory:2020dja,Chagnet:2021uvi} (for the application of similar mathematical structures to a description of other definitions for complexity, see~\cite{Caputa:2021sib,Patramanis:2021lkx}).

\begin{figure}[t!]
\centerline{\includegraphics[scale=.3]{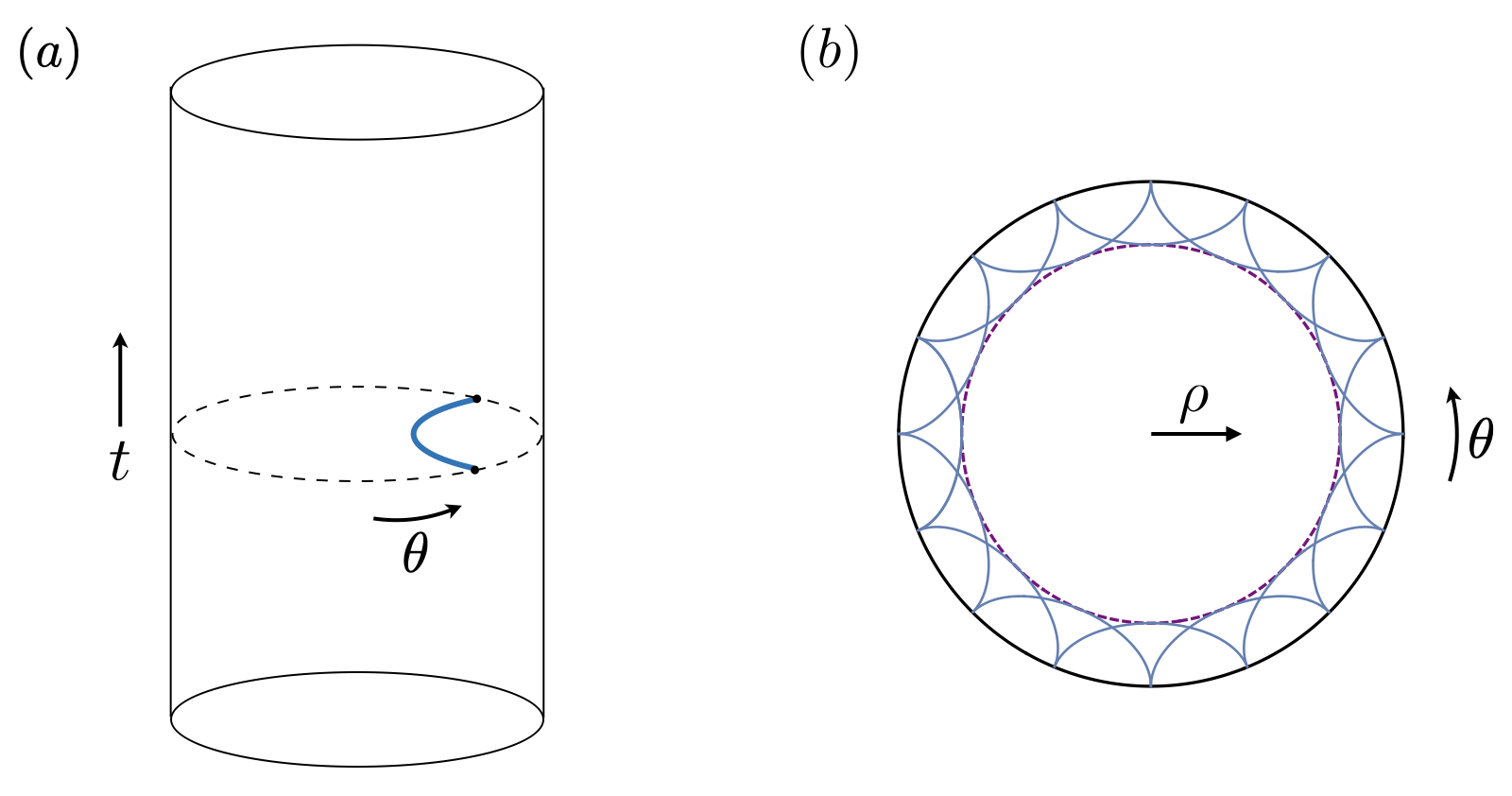}}
\caption{(a) Kinematic space can be defined as the space of pairs of spacelike separated points in a CFT, which are in correspondence with bulk minimal area spacelike geodesics ending on these points. The blue curve is one such geodesic, in the special case that the endpoints lie on the same time slice. (b) The parallel transport of operators in kinematic space can be related to lengths in the bulk AdS spacetime. Depicted here is a constant time slice of anti-de Sitter spacetime. Pairs of points on the boundary define bulk geodesics (blue, solid curves). As the interval position is varied, these trace out an envelope in the bulk (dashed purple circle). The length of this envelope is directly related to the Berry phase associated to the boundary parallel transport of bilocal operators evaluated at the endpoints~\cite{Czech:2017zfq}.}
\label{fig:geodesicplot}
\end{figure}

Another application arises in a subfield of holography known as `kinematic space,' which studies the geometric properties of the space of spacelike pairs of points in a CFT$_d$ and their role in probing the geometry of the bulk anti-de Sitter (AdS) spacetime~\cite{Czech:2015qta,deBoer:2015kda,Czech:2016xec,deBoer:2016pqk}. It was demonstrated that certain bilocal operators in a CFT pick up phases under a parallel transport that displaces the location of the spacelike points where they are evaluated. In the bulk AdS spacetime this was shown to compute the length of a curve traced out by geodesics limiting to these point pairs on the boundary (see Figure~\ref{fig:geodesicplot})~\cite{Czech:2017zfq}. As we show in Appendix~\ref{app:kinematic}, these results for kinematic space can be understood using the language of state-based parallel transport. 

\subsection{Density matrices} \label{sec:density}
Consider a subregion $A$ on a time slice of a CFT. Associated to this region is an algebra of operators $\mathcal A_A$. Assuming some short distance cutoff, the state is described by a reduced density matrix $\rho_A$, obtained from tracing the full state over the complement $\bar{A}$ of $A$. From this we can define the modular Hamiltonian $H_{\rm mod}$ through $\rho_A = e^{- H_{\rm mod}}/(\mbox{tr } e^{- H_{\rm mod}})$. The modular Hamiltonian encodes information about the entanglement properties of the state. It will be formally useful to refer to the `complete' modular Hamiltonian $H_{\rm mod, A} - H_{\rm mod, \bar{A}}$. We will often drop the subscript $A$, and additionally allow the modular Hamiltonian to depend on some parameter $H_{\rm mod}(\lambda)$. This could for instance encode changes in the size of region $A$ as was studied in~\cite{Czech:2017zfq, Czech:2019vih}, or a change of state as we describe in the next section.

The physical data associated to $A$ is not the set of operators in $\mathcal{A}$, but rather their expectation values. As such, there can be symmetries, i.e, transformations which act on the algebra while leaving no imprint on measurable quantities. We define a \emph{modular zero mode} $Q_i$ as a Hermitian operator that commutes with the modular Hamiltonian, \\
\be [Q_i,H_{\rm mod}] = 0~. \ee
The modular zero mode can be exponentiated to the unitary
\be V = e^{-i \sum_i s_i Q_i}~. \label{eq:V}\ee
Under the flow $\mathcal O \rightarrow V^\dagger \mathcal O V$, the expectation values of algebra elements are left unchanged while taking the algebra to itself. The transformation by modular zero modes therefore constitutes a kind of gauge redundancy.

Given an operator, it is often useful to separate the zero mode part out from a contribution that is non-ambiguous. In the finite-dimensional case, we can compute the zero mode contribution by using the projection operator
\be P_0[\mathcal O] = \sum_{E, q_i, q_i'} \ket{E, q_i} \bra{E, q_i} \mathcal O \ket{E, q_i'}\bra{E,q_i'}~, \label{eq:projection} \ee
where $\ket{E,q_i}$ are simultaneous eigenstates of $H_{\rm mod}$ and $Q_i$. Note that later we will be working with an infinite-dimensional algebra, where this formula no longer applies. We will show how to define an appropriate projection relevant for that situation in Section~\ref{sec:transport}.

The zero mode frame redundancy leads to a Berry transport problem for operators. Imagine a process that modifies the algebra $\mathcal{A}_A$ depending on a parameter $\lambda$, for instance by changing the interval $A$ or the state. We start by diagonalizing the modular Hamiltonian,
\be H_{\rm mod} = U^\dagger \Delta U~, \label{eq:HmodinU}\ee
where $\Delta$ is a diagonal matrix of eigenvalues. $H_{\rm mod}, U$ and $\Delta$ are functions of $\lambda$ that vary along the path. Taking the derivative gives the `parallel transport equation,'
\be \dot{H}_{\rm mod} = [\dot{U}^\dagger U, H_{\rm mod}] + U^\dagger \dot{\Delta} U~, \label{eq:paralleltransport}\ee
where $\cdot = \partial_\lambda$. The first term on the right-hand side lies in the image of the adjoint action, $[\cdot, H_{\rm mod}].$ The second term encodes the change of spectrum under the parallel transport. It is a zero mode since it commutes with the modular Hamiltonian, in other words, it lies in the kernel of the adjoint action. We will assume that there is a unique decomposition into the image and kernel of the adjoint action, so that the entire zero mode contribution can be isolated from the second term: $P_0[\dot{H}_{\rm mod}] = U^\dagger \dot{\Delta} U$. For a discussion of subtleties associated with this assumption for the Virasoro algebra, see Appendix~\ref{sec:NonDiagonalization}.

This equation exhibits a redundancy due to the presence of modular zero modes. For instance, the modular Hamiltonian together with Eq.~\eqref{eq:paralleltransport} could be equally well expressed in terms of $U\rightarrow \tilde U = UV$ where $V$ given by Eq.~\eqref{eq:V} is generated by a modular zero mode. Instead of Eq.~\eqref{eq:HmodinU} this gauge choice leads to
\be H_{\rm mod} = V^\dagger U^\dagger \Delta U V~.\ee
A reasonable choice for fixing this ambiguity is to impose that
\be P_0[\partial_\lambda\tilde U^\dagger \tilde U] = 0~.\label{eq:projectionassump}\ee
Since $V$ preserves the zero mode space, $P_0[V^\dagger \dot{U}^\dagger U V] = V^\dagger P_0[\dot{U}^\dagger U]V$ from Eq.~\eqref{eq:projection}. Likewise, $\dot{V^\dagger} V$ is a modular zero mode from Eq.~\eqref{eq:V}, so it projects to itself. Thus, this condition reduces to
\be -V^\dagger \dot{V} + V^\dagger P_0[\dot{U^\dagger}U] V = 0~, \ee
where we have used $\dot{V^\dagger} V = - V^\dagger \dot{V}$ since $V$ is unitary. We therefore obtain a more familiar expression for parallel transport of the operator $V$,
\be \left(\partial_\lambda - \Gamma\right) V = 0~, \ee
where 
\be \Gamma = P_0[\dot{U^\dagger} U] \ee
is a Berry connection that encodes information about how the zero mode frame changes as we vary the modular Hamiltonian. It transforms as $\Gamma \rightarrow V^\dagger \Gamma V - V^\dagger \dot{V}$ under $U\rightarrow U V$. After performing the parallel transport around a closed loop, $\dot{U}^\dagger U$ has a definite value by Eq.~\eqref{eq:projectionassump}. However, $U$ itself may differ by a modular zero mode,
\be U(\lambda_f) = U(\lambda_i) e^{-i \sum_i \kappa_i Q_i}~. \ee
Here, $\lambda_f=\lambda_i$ are the endpoints of a closed path. The coefficients $\kappa_i$ contain information about the loop. 

There is also a curvature, $F$, associated to this parallel transport process. We can evaluate the curvature by performing parallel transport around a small loop. Here, `small' means that we replace the derivatives with infinitesimal transformations. We can think of the operator $S_{\delta \lambda}=\tilde{U}^\dagger \delta_{\lambda}\tilde U$ as a generator of parallel transport. It transforms as a gauge field \be
S_{\delta \lambda} \to V^{\dagger} S_{\delta \lambda} V + V^{\dagger}\delta_{\lambda }V 
\ee 
under a change of modular frame $\tilde{U}\to \tilde{U}V$ and satisfies $P_0[S_{\delta \lambda}]=0$ by Eq.~\eqref{eq:projectionassump}. The curvature $F$ associated to this gauge field is what we call the \emph{modular Berry curvature}. It can be represented in the usual way by performing two consecutive infinitesimal transformations $\lambda_i \to \lambda_i+\delta_1\lambda$, followed by  $\lambda_i+\delta_1\lambda \to \lambda_i+\delta_1\lambda+\delta_2\lambda$. Doing the same with $(1 \leftrightarrow 2)$ and taking the difference gives a closed loop with 
\be\label{eq:curvature2}
F=(1+S_{\delta_2\lambda}(\lambda_i+\delta_1\lambda))(1+S_{\delta_1\lambda}(\lambda_i))-(1 \leftrightarrow 2)~.
\ee
Here, we use that the holonomy operator along the line $[\lambda_i,\lambda_i+\delta\lambda]$ is given by 
\be
\exp \left(\int_{\lambda_i}^{\lambda_i+\delta \lambda} \tilde{U}^{\dagger}\delta_{\lambda}\tilde{U} \right) =1+S_{\delta \lambda}(\lambda_i)~.
\ee
In Appendix~\ref{app:general}, we will derive a general expression for the curvature, Eq.~\eqref{eq:curvature2}, and we apply it in Section~\ref{sec:transport} to the case of state-changing parallel transport. 
\subsubsection{Example: Shape-changing parallel transport}\label{sec:IntervalTransport}

As an example, we will review how this framework for parallel transport of operators can be applied to a parallel transport process of the modular Hamiltonian intervals whose location is varied in the CFT vacuum. This reduces to the study of kinematic space, which we see can also be described using state-based parallel transport in Appendix~\ref{app:kinematic}.

We consider our subregion $A$ to be an interval on a fixed time slice of the CFT with endpoints located at $\theta_L$ and $\theta_R$. Generalizing to subregions with endpoints which are not in the same time slice is straightforward. The modular Hamiltonian associated to $A$ can be written in terms of $\mathfrak{sl}(2,\mathbb{R})$ generators as  
\begin{align}\label{eq:Kplus}
H_{\rm mod} & = s_1 L_1 + s_0 L_0 + s_{-1}L_{-1}~.
\end{align}
Here, we have omitted the  $\bar{L}$ operators for simplicity. The coefficients in Eq.~(\ref{eq:Kplus}) depend on $\theta_L,\theta_R$ and can be determined by requiring that the generator keeps the interval fixed. Explicitly, they are given by 
\be \label{eq:coefficients}
s_1 =  \frac{2\pi \cot \left(\frac{\theta_R -\theta_L}{2}\right)}{e^{i \theta_R} + e^{i \theta_L}}~, ~~~
s_0 = -2\pi \cot\left(\frac{\theta_R - \theta_L}{2}\right)~, ~~~
s_{-1} = \frac{2\pi \cot \left(\frac{\theta_R -\theta_L}{2}\right)}{e^{-i \theta_R} + e^{-i \theta_L}}~. 
\ee
In case of $A$ extending along half the interval, taking for example $\theta_R=-\theta_L=\pi/2$, the modular Hamiltonian can be found from Eq.~\eqref{eq:coefficients} to be $H_{\rm mod}=\pi(L_1+L_{-1})$.

We now construct a one-parameter family of modular Hamiltonians by changing the shape of the interval. The simplest trajectory is given by just changing one of the endpoints, e.g., taking the parameter $\lambda = \theta_L$. The change in modular Hamiltonian is now captured by the parallel transport equation Eq.~\eqref{eq:paralleltransport}, which in this case reads
\be \delta_{\theta_L}H_{\rm mod} = [S_{\delta \theta_L}, H_{\rm mod}]~. \label{eq:shapechanging}\ee
We can solve Eq.~\eqref{eq:shapechanging} for the shape-changing parallel transport operator $S_{\delta \theta_L}$ by first diagonalizing the action of the modular Hamiltonian 
\be\label{eq:eigen}
[H_{\rm mod},V_\mu] = i \mu V_\mu~,
\ee
with $\mu \in {\mathbb{R}}$. It is not difficult to see that the following operators are solutions
\begin{align}
V_{-2\pi}=\partial_{\theta_L}H_{\rm mod}~,\quad V_0=H_{\rm mod}~, \quad V_{2\pi}=\partial_{\theta_R}H_{\rm mod}~,
\end{align}
with $\mu=-2\pi,0,2\pi$ respectively. The operators $V_{2\pi}$ and $V_{-2\pi}$ saturate the modular chaos bound~\cite{DeBoer:2019kdj}. 
Importantly, notice that this class of deformations is characterized by imaginary eigenvalues in Eq.~\eqref{eq:eigen}. The generator of modular parallel transport therefore takes the form
\be
S_{\delta \theta_L} = -\frac{i}{2\pi} \partial_{\theta_L}H_{\rm mod}~.
\ee
For this particular operator Eq.~(\ref{eq:projectionassump}) is automatically satisfied, since it can be written as the commutator of $H_{\rm mod}$. Similarly, one can show that $S_{\delta \theta_R} = \frac{i}{2\pi} \partial_{\theta_R}H_{\rm mod}$. Then, using Eq.~(\ref{eq:curvature2}) one can compute the modular Berry curvature for this shape-changing transport to be 
\be \label{eq:curvature3}
F=[S_{\delta\theta_L},S_{\delta \theta_R}] =-\frac{i}{4\pi}\frac{H_{\rm mod}}{\sin^2\left(\frac{\theta_R - \theta_L}{2}\right)}~.
\ee
In particular, applying the projection $P_0$ to this expression does not change it, as the curvature is proportional to a zero mode. In Appendix \ref{app:kinematic}, we rederive the result in  Eq.~\eqref{eq:curvature3} from the point of view of kinematic space. The curvature, Eq.~\eqref{eq:curvature3}, is simply the volume form on kinematic space.

%% file: Section3bis.tex
\section{State-changing parallel transport}\label{sec:transport}

Let us apply the formalism above to a parallel transport process that modifies not the location of the entangling interval, but rather the state of the system. For definiteness, we work on the AdS$_3$ cylinder with a choice of time slice in the boundary CFT$_2$. 

Consider a change of state by acting by an element $\xi(z)$ of Diff($S^1$), starting from the vacuum of AdS$_3$. The operator that implements this is
\be X_{\xi} = \frac{1}{2\pi i} \oint \xi(z) T(z)\, dz~,\ee
where $T(z)$ is the stress tensor of the boundary CFT. In particular, the diffeomorphism $\xi(z)= z^n$ is implemented by the usual Virasoro mode operator $X_{z^n}=L_{n-1}$. 

Under such a general transformation, the modular Hamiltonian $H_{\rm mod}$ associated to some interval on the boundary transforms as
\be \delta_{\xi} H_{\rm mod} =[X_{\xi},H_{\rm mod}]~. \label{eq:PT}\ee
 Notice that this is just the parallel transport equation, Eq.~\eqref{eq:paralleltransport}, minus the zero mode piece.

Now imagine computing the curvature, Eq.~\eqref{eq:curvature2}, by taking the parallel transport along a small square, i.e., first performing a transformation $\xi_1$ followed by a transformation $\xi_2$, then subtracting the opposite order. The result for the curvature is derived in Appendix~\ref{app:general} and is given by
\be F = P_0([X_{\xi_1},X_{\xi_2}])~, \label{eq:berrycurvature}\ee
where $P_0$ projects to the zero mode of its argument, and the operators $X_{\xi_i}$ are assumed to have no zero modes themselves. We note that while we focus here on CFT$_2$, this is a quite general result that applies to any parallel transport process of the form Eq.~\eqref{eq:PT}. Eq.~\eqref{eq:berrycurvature} together with its application in an explicit example constitute the main results of this section. 

The projection operator in Eq.~\eqref{eq:berrycurvature} is defined by the property that it gives a nonzero answer when evaluated on the modular Hamiltonian (and in general, any other operators that commute with it). Meanwhile, it evaluates to zero on any other operators, which we have assumed take the form $[\cdot, H_{\rm mod}]$ in the decomposition Eq.~\eqref{eq:paralleltransport}. It is possible to construct the projection explicitly in cases where the modular Hamiltonian is known, for instance in our case of CFT$_2$. Let $\theta$ be the spatial boundary coordinate on a constant time slice. The modular Hamiltonian for an interval of angular radius $\alpha$ centered around $\theta=0$ on the cylinder is~\cite{Cardy:2016fqc, Blanco:2013joa}
\be H_{\rm mod} = \int_{-\alpha}^{\alpha} d\theta\, \frac{\cos{\theta}-\cos{\alpha}}{\sin{\alpha}} \,T_{00}(\theta)~. \label{eq:Hmodtheta}\ee
Here, the units are chosen so that the stress energy tensor is dimensionless, $T_{00} \sim -c/12$  in the vacuum on the cylinder, with $T_{00}(\theta)\equiv-(T(\theta)+\overline{T}(\theta))$.

It will be useful to work in planar coordinates. We consider the conformal transformation
\begin{equation}
     z=e^{i\theta}~\label{eq:cylindertoplane}
\end{equation}
to map the cylinder to the plane (with radial ordering). In particular, the interval $[-\alpha,\alpha]$ in the $\theta$-coordinate is mapped to the circle arc with opening angle $2\alpha$ in the $z$-plane. The stress tensor transforms as
\be
T(\theta)=\left(\frac{\partial z}{\partial \theta} \right)^{2}T(z)+\frac{c}{12}\{z,\theta\}~,
\ee
where the Schwarzian derivative is defined by 
\be \{z,\theta\}=\frac{z'''}{z'}-\frac{3}{2} \left(\frac{z''}{z'}\right)^2~.
\ee
Applying the transformation Eq.~\eqref{eq:cylindertoplane}, we find that the modular Hamiltonian on the plane is given by
\be H_{\rm mod} = \frac{1}{i}\oint_{|z|=1} \frac{\frac{1}{2}(1+z^2) - z \cos{\alpha}}{\sin{\alpha}} T(z) \,dz~. \label{eq:Hmodplane1}\ee
Notice that in Eq.~\eqref{eq:Hmodplane1} we have converted to the \emph{complete} modular Hamiltonian by integrating over the full range of coordinates instead of $[-\alpha,\alpha]$.
The reason is that an integration over the full circle allows for an expansion of quantities in terms of Virasoro modes. Moreover, we have conveniently subtracted the vacuum energy of the cylinder in going from Eq. \eqref{eq:Hmodtheta} to Eq. \eqref{eq:Hmodplane1} and only kept the holomorphic part of the stress tensor.

For simplicity, we will take $\alpha = \pi/2$ so that the interval extends along half of the cylinder (from $z=-i$ to $z=i$ in the Euclidean plane). The generalization to intervals with arbitrary $\alpha$ is straightforward. With this convention the modular Hamiltonian simplifies to
\begin{equation}
   H_{\rm mod} =\frac{1}{2 i}\oint (1+z^2)T(z)\,dz~. \label{eq:Hmodplane}
\end{equation}
We can also express this in terms of the Virasoro modes on the plane,
\begin{equation}
    L_n= \frac{1}{2\pi i}\oint z^{n+1}T(z)dz~\label{eq:Ln}~,
\end{equation}
which satisfy the Virasoro algebra 
\be \label{eq:virasoroalgebra}
[L_m,L_n]=(m-n)L_{m+n}+\frac{c}{12}m(m^2-1)\delta_{m+n,0}~.
\ee
Then, Eq.~\eqref{eq:Hmodplane} can be re-expressed as
\begin{equation} \label{eq:HmodVir}
    H_{\rm mod}=\pi (L_{-1}+L_{1})~.
\end{equation}

In the following, it will be useful to write formulae in terms of the diffeomorphism $\xi$ directly, rather than in terms of the corresponding operator $X_{\xi}$. In particular, we identify the modular Hamiltonian $H_{\rm mod}$ with the vector field $\xi(z)=\pi(1+z^2)$, as follows from Eq.~\eqref{eq:Hmodplane}. Moreover, if we take an operator of the form
\be X_{\xi} = \frac{1}{2\pi i} \oint \xi(z) T(z)\, dz~,\label{eq:Xchi}\ee
the commutator with $H_{\rm mod}$ can also be expressed in $\xi$ directly. Using Eqs.~\eqref{eq:Hmodplane} and \eqref{eq:Xchi}, applying the OPE
\be T(w) T(z) = \frac{c/2}{(z-w)^4} + \frac{2T(w)}{(z-w)^2} + \frac{\partial T(w)}{z-w} + ...\ee
and integrating by parts we find
\be 
[H_{\rm mod},X_{\xi}] = \frac{1}{2i} \oint \left[2z \xi(z)-(1+z^2)\xi'(z)\right]T(z)\, dz~.\label{eq:nonzeromode1}
\ee 
Applying several integration by parts directly onto Eq.~\eqref{eq:Hmodplane}, the term proportional to the central charge identically vanishes in this case.

To implement Eq. \eqref{eq:berrycurvature} for the modular Berry curvature one needs to define the operator $P_0$ which projects onto the zero mode. Following the general prescription in Section \ref{sec:density}, one would like to decompose an arbitrary operator $X$ into the image and the kernel of the adjoint action of $H_{\rm mod}$, 
\be \label{eq:decomposition1}
X=\kappa H_{\rm mod}+[H_{\rm mod},Y]~,
\ee
where $\kappa$ is the zero mode that needs to be extracted. 
However, it turns out that there is a subtlety associated with the above decomposition in the case of the Virasoro algebra. In general, there are operators which are neither in the kernel, nor in the image of the adjoint action\footnote{For finite-dimensional vector spaces this is not the case if the kernel and image are disjoint, as follows from a simple dimension counting. In the infinite-dimensional set-up the situation is more complicated, e.g., one can write down linear maps which are injective but not surjective.}, which leads to an ambiguity in the definition of the zero mode projection $P_0$. We refer to Appendix \ref{sec:NonDiagonalization} for a discussion of these issues in the case of the Virasoro algebra. 
For this reason, we will consider a different class of transformations, i.e., those which diagonalize the adjoint action of the modular Hamiltonian $H_{\rm mod}$ (see \cite{Das:2020goe} where a similar diagonalization in terms of so-called modular eigenmodes was considered). Therefore, we start from the eigenvalue equation
\begin{equation} \label{eq:eigenvalueeq}
[H_{\rm mod},X_{\lambda}]=\lambda X_{\lambda}~,
\end{equation}
where we have used the short-hand notation $X_{\lambda}\equiv X_{\xi_{\lambda}}$ for the operator associated to the transformation $\xi_{\lambda}$. Using Eq.~\eqref{eq:nonzeromode1} it is not difficult to see that Eq.~\eqref{eq:eigenvalueeq} is solved by
\begin{equation} \label{eq:eigenfunction}
\xi_{\lambda}(z)=\pi(1+z^2)\left(\frac{1-iz}{z-i} \right)^{-i\lambda/2\pi}~.
\end{equation}
In particular, we see that the operator with eigenvalue zero, $\lambda =0$, is the modular Hamiltonian itself, as one would expect from Eq.~\eqref{eq:eigenvalueeq}. Notice that the solutions in Eq.~\eqref{eq:eigenfunction} go to zero at the endpoints of the interval:
\begin{equation}\label{eq:vanishing}
\xi_{\lambda}(z)\to 0 \hspace{10pt} {\rm as} \hspace{10pt} z\to \pm i~.	
\end{equation}	
The eigenfunctions of $H_{\rm mod}$ therefore correspond to the transformations which change the state, but not the location of the boundary interval. They are not analytic at $z=\pm i$,\footnote{Note that due to Eq.~\eqref{eq:vanishing}, it is valid to apply a single integration by parts. Thus, Eq.~\eqref{eq:nonzeromode1} is maintained.} so strictly speaking they are not part of the Virasoro algebra (defined in the usual way as the space of smooth vector fields on the circle). However, they seem to be the natural transformations to consider in this context. We will refer to them as \emph{state-changing} transformations as opposed to the shape-changing transformations in Section \ref{sec:IntervalTransport}. 

From Eq.~\eqref{eq:eigenvalueeq} combined with the Jacobi identity, these eigenfunctions form an algebra with commutation relations 
\begin{equation} \label{eq:algebra}
[X_{\lambda},X_{\mu}]=(\lambda-\mu)X_{\lambda+\mu}~,
\end{equation}
which defines a continuous version of the Virasoro algebra\footnote{A Virasoro algebra with continuous index also appears in the context of the so-called dipolar quantization of 2d CFT \cite{Ishibashi:2015jba, Ishibashi:2016bey} which is related to the sine-square deformation~\cite{Gendiar:2008udd,Katsura:2011ss}, as well as in the study of non-equilibrium flows in CFT~\cite{Doyon:2013paa}.} with generators $X_{\lambda}$ labeled by a continuous parameter $\lambda \in \mathbb{R}$. Note that in the following we are leaving out the central extension (so strictly speaking we are working with a continuous version of the Witt algebra). We will return to discuss how to include the central extension in Section~\ref{sec:extension}.

It is natural to define the transformations in Eq.~\eqref{eq:eigenfunction} to have support only on the subregion $A$. In the case at hand, this makes all the contour integrals collapse to integrals over the semicircle from $-i$ to $i$, e.g.,~the $\lambda=0$ eigenfunction does not correspond to the \emph{complete} modular Hamiltonian, but simply to the half-sided one. The state-changing vector fields, which might look unfamiliar in terms of the $z$-coordinate, take a more familiar form when we map the entanglement wedge to a hyperbolic black hole geometry using \cite{Casini:2011kv}.

This can be seen in the following way. Starting with the boundary CFT$_d$ on the Euclidean cylinder $\mathbb{R} \times S^{d-1}$ with metric
\be\label{cylinder}
ds^2 = dt_E^2 + d \theta^2 + \sin^2 \theta ~ d\Omega_{d-2}^2~,
\ee
we consider a fixed sphere at $t_E=0$, $\theta = \theta_0$. We can apply the following conformal transformation considered in~\cite{Casini:2011kv}: 
\bea \label{CHM}
\tanh t_E &=& \frac{\sin \theta_0 \sin \tau }{\cosh u + \cos \theta_0 \cos \tau}~, \nonumber \\  
 \tan \theta &=& \frac{\sin \theta_0 \sinh u}{\cos \theta_0 \cosh u + \cos \tau}~, 
\eea
which conformally maps the causal development of the sphere to the hyperbolic geometry $\mathbb{R}\times\mathbb{H}^{d-1}$ given by
\be
ds^2 = \Omega^2 \left(d\tau^2 + du^2 + \sinh^2 u ~ d \Omega^2_{d-2} \right)~,\\
\ee
with conformal factor
\be
\Omega^2=\frac{\sin^2\theta_0}{(\cosh u +\cos \theta_0\cos\tau)^2-\sin^2\theta_0\sin^2\tau}~.
\ee

Taking $d=2$ and $\theta_0 = \pi /2$ for the half interval entangling surface, the transformation Eq.~(\ref{CHM}) at the $\tau=0$ (or equivalently $t_E=0$) time slice reduces simply to 
\be
\tan \theta =\sinh u~.
\ee
Written in terms of the coordinate $z=e^{i\theta}$ this leads to
\begin{equation} \label{eq:coordtrans}
e^{u}=\frac{1-iz}{z-i}~.	
\end{equation}
Recall that the boundary region $A$ corresponds to $|z|=1$ and $-\pi/2\leq \arg(z)\leq\pi/2$ in the plane, so it is mapped to $u\in \mathbb{R}$. Moreover, the components of the vector field transform according to
\be \label{eq:vectorfieldtrans}
\xi_{\lambda}(z)\frac{\partial}{\partial z}=	\xi_{\lambda}(u)\frac{\partial}{\partial u}~
\ee
with
\be\label{eq:uzcoordtransf}
du =-2i\frac{dz}{1+z^2}~,
\ee
so that the transformations take the simple form
\begin{equation} \label{eq:uvariable}
\xi_{\lambda}(u)=-2\pi i\,e^{-i\lambda u/2\pi}~.	
\end{equation}
Hence, we find that the state-changing transformations, when written in terms of the $u$-variable, are simply plane wave solutions with frequency $\lambda/2\pi$ in this black hole background. Therefore, they are natural objects to consider in this geometry. 

We can reintroduce both the right- and the left-movers by replacing $u\to u+i\tau$ in Eq.~\eqref{eq:coordtrans}, so that $z$ is allowed to take values in the half plane $\mathrm{Re}\, z\geq 0$ (the radial direction in the $z$-plane corresponds to time evolution in $\tau$). Eq.~\eqref{eq:vectorfieldtrans} is therefore modified according to 
\be
\xi_{\lambda}(z)\frac{\partial}{\partial z}=\xi_{\lambda}(u+i\tau)\left(\frac{\partial}{\partial u}-i\frac{\partial}{\partial \tau}\right)~, \quad \xi_{\lambda}(\bar{z})\frac{\partial}{\partial \bar{z}}=-\xi_{\lambda}(-u+i\tau)\left(\frac{\partial}{\partial u}+i\frac{\partial}{\partial \tau}\right)~.
\ee
By setting $\lambda=0$ and adding the right- and left-moving contributions, we see that the modular Hamiltonian indeed acts by time  translation in the black hole geometry:
\begin{equation}
H_{\rm mod}\sim \frac{\partial }{\partial \tau}~.
\end{equation}

Working in the algebra associated to the eigenfunctions of $H_{\rm mod}$, we do have a unique decomposition of the form Eq.~\eqref{eq:decomposition1}: one simply decomposes an arbitrary operator into eigenoperators, which have either $\lambda=0$ or $\lambda\neq 0$. Given such a decomposition it is easy to write down an operation which extracts the zero mode $\kappa$, namely a linear functional $P_0$ which satisfies\footnote{For technical reasons we set $P_0(H_{\rm mod})\sim \delta(0)$, instead of $P_0(H_{\rm mod})\sim 1$ as one might have naively expected. This results from the plane-wave normalizability of the eigenfunctions, Eq.~\eqref{eq:uvariable}. It ensures the modular Berry curvature is finite when evaluated on wave packets in Section \ref{sec:example}.}
\begin{equation} \label{eq:projconditions}
P_0(H_{\rm mod})\sim \delta(0)~, \hspace{10pt} P_0([H_{\rm mod},Y])=0~.
\end{equation}  
In the $u$-coordinate such a functional can be written as
\begin{equation} \label{eq:projectionoperator}
P_0(X_{\xi})=\lim_{\Lambda\to \infty}\frac{i}{2\pi}\int_{-\Lambda}^{\Lambda}\xi(u)\,du~.	
\end{equation}
Using the coordinate change Eq.~\eqref{eq:uzcoordtransf}, we can represent the projection in the $z$-coordinate as
\begin{equation}\label{eq:projectioninz}
P_0(X_\xi) =\lim_{\Lambda\to \infty}\frac{i}{2\pi}\int_{-\Lambda}^{\Lambda}\xi(u)\, du = \frac{1}{\pi}\int_{-i}^{i}\frac{\xi(z)}{(1+z^2)^2}\,dz~.
\end{equation}
When applied to the eigenfunctions of $H_{\rm mod}$ the projection becomes
\begin{equation}
P_0(X_{\lambda})=\lim_{\Lambda\to \infty}2\pi\int_{-\Lambda}^{\Lambda} e^{i\lambda u}\,du=4\pi^2\delta(\lambda)~,
\end{equation}
which is a standard representation of the Dirac delta function. To show that $P_0$ vanishes on commutators of the form $[H_{\rm mod},Y]$, it suffices to remark that one can take $Y$ to satisfy $[H_{\rm mod},Y]=\lambda Y$ with $\lambda\neq 0$ without loss of generality. This shows that Eq.~\eqref{eq:projectionoperator} defines a good projection operator in the sense of Eq.~\eqref{eq:projconditions}. Unlike for the case of the ordinary Virasoro algebra treated in Section~\ref{app:nondiag}, there is no ambiguity in the resulting projection.

\subsection{Example} \label{sec:example}
We now have all the ingredients to compute the curvature in an explicit example. We consider a general perturbation of the form
\begin{align}
	z'=z+\epsilon \, \xi(z)+\mathcal{O}(\epsilon^2)~, \label{eq:transf}
\end{align}
where $\xi(z)$ is a wave packet
\be
\xi(z)=\frac{1}{2\pi}\int_{-\infty}^{\infty} c(\lambda) \xi_{\lambda}(z)\, d\lambda~,
\ee
with $\xi_{\lambda}(z)$ defined in Eq.~\eqref{eq:eigenfunction}. We start by obtaining the correction to the transformed modular Hamiltonian upon acting with Eq.~\eqref{eq:transf}. Let us expand both the modular Hamiltonian and the parallel transport operator to first order in the small parameter $\epsilon$:
\begin{equation}
	H'_{\rm mod} =H^{(0)}+\epsilon\, H^{(1)}+\mathcal{O}(\epsilon^2)~, \indent S = S^{(0)} + \epsilon\, S^{(1)} + \mathcal{O}(\epsilon^2)~.
\end{equation}
Using that $z=z'-\epsilon \xi(z')+\mathcal{O}(\epsilon^2)$, one can expand the transformed $H_{\rm mod}$ to order $\mathcal{O}(\epsilon^2)$.
One finds that $H^{(0)}=H_{\rm mod}$ is the original modular Hamiltonian, while the correction is given by
\begin{equation} \label{eq:H1}
	H^{(1)}=-\frac{1}{2i}\oint \left[2z\xi(z)-(1+z^2) \xi'(z)\right] T(z)\, dz~.
\end{equation} 
Here, 
we have neglected the Schwarzian contribution for simplicity. It will be treated separately in Section \ref{sec:extension}. We now expand the parallel transport equation
\begin{equation}
\delta H_{\rm mod}=[S,H_{\rm mod}]
\end{equation}
to first order in $\epsilon$. This gives two separate equations: 
\be
0= [S^{(0)}, H^{(0)}]~, \quad 
H^{(1)} = [S^{(0)},H^{(1)}]+[S^{(1)},H^{(0)}]\label{eq:transportfirst}~.
\ee
Solving  Eq.~\eqref{eq:transportfirst} for the correction $S^{(1)}$ to the parallel transport operator gives the solution
\begin{equation}
S^{(0)}=0~, \indent S^{(1)}=X_{\xi}~.
\end{equation}
Both $S^{(0)}$ and $S^{(1)}$ are defined up to a zero mode, meaning that one can add to it an extra operator $Q$ for which $[Q,H_{\rm mod}]=0$ (e.g., the modular Hamiltonian itself) and the parallel transport equation would still be satisfied.

To compute the curvature we need to consider two different parallel transport operators $S_1$ and $S_2$ which we take to be defined according to the transformations
\be
\xi_1(z)=\frac{1}{2\pi}\int_{-\infty}^{\infty} c_1(\lambda) \xi_{\lambda}(z)\, d\lambda~, \quad \xi_2(z)=\frac{1}{2\pi}\int_{-\infty}^{\infty} c_2(\lambda) \xi_{\lambda}(z)\, d\lambda~,
\ee
respectively. After projecting out their zero modes, we take the commutator and project to the zero modes again to obtain the value of the curvature component. Therefore, we need to compute
\begin{equation} \label{eq:comm}
[S^{(1)}_1-\kappa_1H^{(0)},S^{(1)}_2-\kappa_2H^{(0)}]~,
\end{equation}
where $\kappa_i=P_0(S_i)$, is the zero mode coefficient of the parallel transport operator $S_i$. We can split Eq.~\eqref{eq:comm} into terms that we can treat separately. Notice that the term proportional to $[H^{(0)},H^{(0)}]$ is zero and can be removed. Moreover, the definition of the projection operator immediately implies
\begin{equation}
P_0([S^{(1)}_1,H^{(0)}])=P_0([S^{(1)}_2,H^{(0)}])=0~.
\end{equation} 
To evaluate the last commutator we use the commutation relations in Eq.~\eqref{eq:algebra} to obtain 
\begin{equation}
[S^{(1)}_1,S^{(1)}_2]=\frac{1}{4\pi^2}\int_{-\infty}^{\infty}\int_{-\infty}^{\infty} (\lambda_1-\lambda_2)c_1(\lambda_1) c_2(\lambda_2) X_{\lambda_1+\lambda_2}\, d\lambda_1d\lambda_2~.
\end{equation}
Applying the projection operator sets $\lambda_1=-\lambda_2$, so that we find
\begin{equation}
P_0([S^{(1)}_1,S^{(1)}_2])=2\int_{-\infty}^{\infty}\lambda\, c_1(\lambda)c_2(-\lambda)\, d\lambda~.
\end{equation} 
Therefore, the final result for the modular Berry curvature associated to the state-changing transport problem is given by
\begin{equation}\label{eq:curvature}
F=2\int_{-\infty}^{\infty}\lambda\, c_1(\lambda)c_2(-\lambda)\, d\lambda~.
\end{equation}
Note that the curvature appropriately vanishes when two perturbations lie along the same direction, $c_1(\lambda)=c_2(\lambda)$. If we take the modes to be peaked at the eigenfunctions $\xi_{\lambda_i}(z)$ themselves, $c_i(\lambda)=\delta(\lambda-\lambda_i)$, the above formula reduces to
\begin{equation}
F=(\lambda_1-\lambda_2)\delta(\lambda_1+\lambda_2)~, 
\end{equation}
which is a local formula in terms of the parameters $\lambda_i$.

\subsection{Lie algebra} \label{sec:algebra}

To diagonalize the adjoint action, we saw that we must work with a continuous version of the Virasoro algebra. Viewed in terms of vector fields on the circle, we must consider non-smooth vector fields on the circle, Eq.~\eqref{eq:eigenfunction}, which have support only along the interval. When mapped to the real line, these are just plane waves, Eq.~\eqref{eq:uvariable}. In the last section, we performed parallel transport using wave packets constructed out of these eigenfunctions. In terms of the coordinates on the real line,
\be\label{eq:wavepacket2}
\xi(u)=\frac{1}{2\pi}\int_{-\infty}^{\infty} c(\lambda) \xi_\lambda(u) d\lambda~.
\ee
Now we would like to be more precise about the sense in which the corresponding vector fields form a Lie algebra. This amounts to imposing extra conditions on $c(\lambda)$ for these to form a closed algebra, along with any other desirable properties.

The simplest choice would be to demand that the $\xi(u)$ be smooth. Then, since the smoothness of functions is preserved under pointwise multiplication, the corresponding vector fields $\xi(u)\partial_u$ will form a closed algebra. However, an arbitrary $\xi(u)$ will not necessarily have finite zero mode projection, nor will there necessarily exist a natural definition for a dual space. To define sensible wave packets we will impose two additional requirements:
\begin{itemize}
    \item There is a notion of Fourier transform that maps the space to itself,
    \item The $\xi(u)$ are integrable. This means that the projection, Eq.~\eqref{eq:projectionoperator}, is finite, and this property is preserved under commutation of the vector fields $\xi(u)\partial_u$~. It also allows us to define a dual space in terms of distributions.
\end{itemize}

To accomplish this, it is convenient to work with wave packets $\xi(u)$ that are \emph{Schwartz functions}. These are smooth, bounded functions whose derivatives are also all bounded: $|u^\alpha\partial^\beta \xi(u)| < \infty$ for all $\alpha,\beta>0$. In other words, they rapidly go to zero as $u\rightarrow\pm \infty$, faster than any reciprocal power of $u$. This definition excludes for example polynomials, but includes polynomials weighted by an exponential $e^{-c |u|^2}$ for $c\in \mathbb{R}$. By the Leibniz rule, the Schwartz space $\mathcal{S}$ is closed under pointwise multiplication, thus the corresponding  vector fields  form a closed Lie algebra. We denote $\mathcal{S}$ for the space of Schwartz functions and $\mathfrak{s}$ for the corresponding algebra of vector fields.

Since these functions are integrable, it is natural to define a dual space $\mathcal{S}'$ consisting of linear functionals $T:\mathcal{S}\rightarrow \mathbb{C}$, in terms of distributions:
\be T[\xi(u)] = \int_{-\infty}^\infty \xi(u) T(u)du~. \ee

A pairing between Schwartz functions and dual elements can be defined from this as $\left<T, \xi\right> \equiv T[\xi(u)]$. Likewise, there is also a dual space $\mathfrak{s}^*$ consisting of linear functionals on $\mathfrak{s}$, the algebra of vector fields. This is inherited from the dual space $\mathcal{S}'$, i.e., it consists of the space of distributions evaluated on Schwartz functions. There is a pairing $\left<\cdot,\cdot\right>$ between $\mathfrak{s}$ and $\mathfrak{s}^*$ which descends from the pairing on $\mathcal{S}$ and $\mathcal{S}'$.

Notice that, evaluated on the wave packets Eq.~\eqref{eq:wavepacket2}, the projection operator Eq.~\eqref{eq:projectionoperator}
\be P_0: \xi(u) \mapsto 2\pi c(0)~ \label{eq:projectionoperator2}\ee
is a linear functional, and thus it is an element of the dual space. The pairing is given by $\left<P_0, \xi\right> = P_0(\xi) = 2\pi c(0)$.

In the coordinates on the circle, recall that this dual element can be expressed from Eq.~\eqref{eq:projectioninz} as
\be P_0: \xi(z) \mapsto \frac{1}{\pi} \int dz \frac{\xi(z)}{(1+z^2)^2}~. \label{eq:projectioninz2}\ee
Notice that this dual element is not a smooth quadratic form on the circle as is typically considered in treatments of the dual space of the Virasoro algebra, but rather a more general distribution that involves singularities at $z=\pm i$\footnote{In the usual discussion of the Virasoro algebra the dual space is identified with the space of smooth quadratic differentials. Formally, one could argue that distributions such as $\delta(z-z_0)$ and $\delta'(z-z_0)$ are also part of some suitably defined notion of the dual space. Indeed, they define linear functionals
\be
\xi\mapsto \xi(z_0)~,\hspace{10pt} \xi\mapsto -\xi'(z_0)~,
\ee
which evaluate a function (or its derivative) at some point $z_0$. The projection operator $P_0$ in Eq. \eqref{eq:projectioninz2}, when integrated over the full circle and properly regularized, can be regarded in this fashion. See Appendix \ref{sec:NonDiagonalization} for more details, for example, Eqs. \eqref{eq:residue} - \eqref{eq:projop2}.}. A standard definition of the dual space is an attempt to get a space that is roughly the same size as the algebra itself. For infinite-dimensional spaces the formal dual is much larger and one needs some additional structure, e.g., that of a Hilbert space, to limit it.  

We emphasize that there is considerable freedom in these definitions.  A different choice would amount to taking a different set-up for varying the state in the parallel transport process. Our definitions allow us to perform parallel transport using wavefunctions that are `physical' in the sense of being Fourier transformable and integrable. The existence of a natural dual space also allows for contact with a geometrical picture in terms of coadjoint orbits, which we describe in the next section.

\subsection{Central extension}\label{sec:extension}
We have so far only considered changing the state with a transformation of the circle. When the transformations are diffeomorphisms on the circle, the group $\rm{Diff}(S^1)$ gets centrally extended to the full Virasoro group, $\rm{Diff}(S^1)\times \mathbb{R}$. Here we are considering a continuous version of the Virasoro generated by the transformations, Eq.~\eqref{eq:eigenfunction}. For the central extension, we proceed in direct analogy with the Virasoro case. In the following, the vector fields $\xi(z)$ should be understood to have non-zero support only between $z=\pm i$, so that this is the only part of the integral over the full circle that contributes.

We consider pairs $(\xi,\alpha)$, where $\xi$ is a vector field of the form Eq.~\eqref{eq:eigenfunction}, which diagonalizes the adjoint action, and $\alpha\in \mathbb{R}$. The Lie bracket is defined as \be  \label{eq:Liebracket}
\left[(\xi,\alpha),(\chi,\beta)\right]=\left(-[\xi,\chi],-\frac{1}{48\pi}\oint dz\,(\xi(z)\chi'''(z)-\xi'''(z)\chi(z))\right)~,
\ee
where $[\xi,\chi]:=\xi\chi'-\chi\xi'$ is the commutator of vector fields. This is identical to the commutators for the Virasoro algebra, with the only difference being that we integrate only over half the circle, and also consider transformations $\xi$ which are not smooth at the endpoints. In terms of the operators $X_\lambda$, this extends the algebra in Eq.~\eqref{eq:algebra} to 
\begin{equation} [\bar{X}_{\bar\lambda},\bar{X}_{\bar \mu}]=(\bar{\lambda}-\bar{\mu})\bar{X}_{\bar{\lambda}+\bar{\mu}}+\frac{c}{12}\bar{\lambda}(\bar{\lambda}^2+1)\delta(\bar{\lambda}+\bar{\mu})~.
\end{equation} 
where we have defined rescaled barred variables through $X_\lambda = -2\pi \bar{X}_\lambda, \lambda= -2\pi\bar{\lambda}$ to bring this to a form that more closely resembles the usual Virasoro algebra with discrete labels.
 
One often introduces a new generator, denoted by $c$, which commutes with all other elements in the algebra, to write
\be
(\xi,\alpha)=\xi(z) \partial_z- i\alpha c~.
\ee
By definition, the central element $c$ commutes with $H_{\rm mod}$, i.e., $[H_{\rm mod},c]=0$. Therefore, we can think about the central element as another zero mode in the parallel transport problem.

Luckily, the situation for the central element is simpler than for the modular Hamiltonian itself. From the form of $H_{\rm mod}$, Eq. \eqref{eq:HmodVir}, and the algebra, Eq. \eqref{eq:Liebracket}, we see that the central element $c$ does not appear in commutators of the form $[H_{\rm mod},X]$. Therefore, the projection onto the coefficient of $c$ is simply given by the linear functional
\be 
(\xi,\alpha)\to \alpha~.
\ee
One way to include the information of the central term is to make the Berry curvature give a $U(1)\times U(1)$-valued number (organized in terms of an extra element which we take to be $c$). More precisely, we define the zero mode projection operator $P^c_0$, which depends on $c$, by
\be
P^c_0((X_{\xi},\alpha))=P_0(X_{\xi})- \alpha c~.
\ee
The first term is the usual zero mode, while the second term keeps track of the central zero mode. It is easy to see how the result for the Berry curvature gets modified. Using  Eq. \eqref{eq:berrycurvature} with $P_0^c$ instead of $P_0$, we see that the formula for the Berry curvature is given by
\be \label{eq:berrycurvaturec}
F=P_0([X_{\xi_1},X_{\xi_2}])+\frac{c}{48\pi}\oint dz\, \left(\xi_1(z)\xi_2'''(z)-\xi_1'''(z)\xi_2(z)\right)~.
\ee

As a consistency check, we can go back to our example in Section \ref{sec:example} and consider the contribution from the Schwarzian term in Eq.~(\ref{eq:H1}). Expanding the parallel transport equation, we need to solve
\begin{equation}
\label{eq:transportfirst2}
H^{(1)}= [S^{(1)}, H^{(0)}]~,
\end{equation}
where the change in the modular Hamiltonian due to the Schwarzian derivative to first order is given by
\be\label{eq:Hschwartz}
H^{(1)}_{\rm Schw} = \frac{c}{24i} \oint dz (1+z^2) \xi'''(z)~,
\ee 
having used that $\{z',z\}=\epsilon \xi''' + {\cal{O}}(\epsilon^2)$. On the full circle, applying three integration by parts, this is just $H^{(1)}_{\rm Schw}=0$ (equivalently, no diffeomorphism has $\xi'''=z^{-1}$ or $z^{-3}$ which would give a pole). The situation is a bit more subtle on the half circle, since due to non-differentiability at the endpoints it is no longer valid to apply integration by parts multiple times. However, it is still the case that none of the eigenfunctions, Eq.~\eqref{eq:eigenfunction}, have $\xi'''=z^{-1}$ or $z^{-3}$, and so the Schwarzian contribution vanishes. Thus, in either case the solution to Eq.~(\ref{eq:transportfirst2}) with the new Lie bracket Eq.~\eqref{eq:Liebracket} is still given by $S^{(1)}=X_{\xi}~$.
The extra contribution to the commutator $[S^{(1)}_1,S^{(1)}_2]$ due to the central charge is indeed given by Eq.~\eqref{eq:berrycurvaturec}. 

Note while it is not possible to apply integration by parts multiple times on Eq.~\eqref{eq:Hschwartz} for the half circle, we have \emph{defined} the central extension as the version that obeys integration by parts three times. This is because we have chosen the antisymmetric combination for the central charge part in Eq.~\eqref{eq:Liebracket}. As a result, our bracket respects the properties of the commutator, $[X_{\xi}, X_{\chi}] = - [X_{\chi},X_{\xi}]$. Likewise, one can check that the Jacobi identity is satisfied. Given elements $(\xi,\alpha),(\chi,\beta),(\rho,\gamma)$ which satisfy the algebra Eq.~\eqref{eq:Liebracket}, we have
\begin{gather} [(\xi,\alpha),[(\chi,\beta),(\rho,\gamma)]]+[(\chi,\beta),[(\rho,\gamma),(\xi,\alpha)]]+[(\rho,\gamma),[(\xi,\alpha),(\chi,\beta)]]\nonumber \\
= \left(0,-\frac{1}{48\pi}\oint \left( [\chi,\rho]\,\xi^{(3)} + [\rho,\xi]\,\chi^{(3)} + [\xi,\chi]\,\rho^{(3)}\right)\right)~.
\end{gather}
We can see this is identically zero by integrating each term by parts once onto the commutator, which vanishes at the interval endpoints by Eq.~\eqref{eq:vanishing} so that there is no boundary contribution. These properties are sufficient to ensure the consistency of the central extension.

%% file: coadjoint.tex
\section{Coadjoint orbit interpretation}\label{sec:coadjoint}

Various versions of the parallel transport problem we consider exhibit connections to the geometry of symplectic manifolds known as coadjoint orbits. For the state-based parallel transport summarized in Section~\ref{sec:states} applied to the Virasoro algebra, connections to coadjoint orbits were described in~\cite{Oblak:2017ect}. In Appendix~\ref{app:kinematic}, we additionally explain how to use state-based parallel transport to obtain coadjoint orbits of $SO(2,1)$, which describe kinematic space~\cite{Penna:2018xqq}. We will begin by reviewing the notion of coadjoint orbits, and then we explain how our operator-based parallel transport can be related to the geometry of orbits.

Consider a Lie group $G$ with Lie algebra $\mathfrak{g}$. Let $\mathfrak{g}^*$ be the dual space, i.e., the space of linear maps $T:\mathfrak{g}\rightarrow \mathbb{C}$. This defines an invariant pairing $\left<T,X\right> \equiv T(X)$ for $X\in \mathfrak{g}, T\in \mathfrak{g}^*$. 
The group $G$ acts on the algebra $\mathfrak{g}$ through the adjoint action,
\be \mbox{Ad}_g(X) = \left.\frac{d}{d\lambda}\left(g e^{\lambda X} g^{-1}\right)\right|_{\lambda=0}~, \indent g\in G, \, X\in \mathfrak{g}~.\label{eq:Ad}\ee
For matrix groups such as $SO(2,1)$, which we consider in Appendix~\ref{app:kinematic}, Eq.~\eqref{eq:Ad} is just $\mbox{Ad}_g(X) = g X g^{-1}$.

The adjoint action of the algebra on itself can be defined from this as
\be \mbox{ad}_X(Y) = \frac{d}{d\rho} \left.(\mbox{Ad}_{e^{\rho X}}(Y))\right|_{\rho=0} = [X,Y]~, \indent X,Y\in \mathfrak{g}~.\ee

The adjoint action descends to an action on the dual space. This \emph{coadjoint action} $\mbox{ad}_X^*$ on $\mathfrak{g}^*$ is defined implicitly through
\be \label{eq:mcDef} \left<\mbox{ad}_X^* z, Y\right> = \left<z, \mbox{ad}_X Y\right>~, \indent z\in \mathfrak{g}^*, \, X,Y\in \mathfrak{g}~. \ee
For a given $T\in \mathfrak{g}^*$, the orbit $\mathcal{O}_T = \{\mbox{ad}_X^*(T)\, | \, X\in \mathfrak{g} \}$ generated by the coadjoint action is known as a \emph{coadjoint orbit}. 

Let $x_1,x_2$ be coadjoint vectors tangent to the orbit $\mathcal{O}_T$, and let $X_1, X_2$ be the adjoint vectors that are dual to these through the invariant pairing. Then, the \emph{Kirillov-Kostant symplectic form} associated to this orbit is~\cite{Witten:1987ty, alekseev1988, kirillov2004, Alekseev:2018pbv}
\be \omega(x_1, x_2) = \left<T,[X_1,X_2]\right>~. \ee
This is manifestly anti-symmetric and $G$-invariant. It is also closed and nondegenerate~\cite{Witten:1987ty}, and hence it defines a symplectic structure on $\mathcal{O}_T$. Thus, coadjoint orbits are naturally symplectic manifolds. For matrix groups, the algebra and dual space are isomorphic through the Cartan-Killing form, which is non-degenerate in this case. It suffices to consider an orbit of the adjoint action, and these generate symplectic manifolds. This is the setting of Appendix~\ref{app:kinematic}. We emphasize that in the general case this is not true and one must work in the dual space.

It will be useful to review the case of the Virasoro group, along with a suitable generalization given by the algebra described in Sections~\ref{sec:algebra} and~\ref{sec:extension} that applies to our case of interest. Recall that the Virasoro group consists of $\rm{Diff}(S^1)$ together with its central extension, $\widehat{\rm{Diff}(S^1)}=\rm{Diff}(S^1)\times \mathbb{R}$. For our problem, we are considering a continuous version of the ordinary Virasoro algebra, with a central extension described in Section~\ref{sec:extension}. In either case, the formulae will be the same, with the difference that in the second scenario the vector fields $\xi$ should be understood to be non-differentiable at the interval endpoints, with vanishing support outside the interval. Thus, in the latter case all integrals should be understood to cover only the range of the interval rather than the full circle.

For either algebra we consider elements $\xi(z)\partial_z - i \alpha c$ where $\xi(z) \partial_z$ is a vector field on the circle (smooth for Virasoro, and of the form Eq.~\eqref{eq:eigenfunction} for its generalization) and $\alpha\in \mathbb{R}$ is a parameter for the central extension, generated by the algebra element $c$. The only non-trivial commutators are
\be \left[\xi_1(z)\partial_z, \xi_2(z)\partial_z\right] = -(\xi_1 \xi_2'-\xi_1' \xi_2)\partial_z + \frac{i c}{48\pi}\oint dz \, (\xi_1 \xi_2''' - \xi_1''' \xi_2)~.\ee
In the Virasoro case, using $L_n =z^{n+1} \partial_z$ the bracket Eq.~\eqref{eq:Liebracket} indeed leads to the usual form of the Virasoro algebra, Eq.~\eqref{eq:virasoroalgebra}. 

For both algebras we can define a pairing between an adjoint vector $(\xi, \alpha)$ and a coadjoint vector $(T,\beta)$ given by
\be \left<(T,\beta),(\xi,\alpha)\right> = -\left[\oint dz \, T(z)\xi(z) + \alpha \beta\right].\ee

Now consider algebra elements $X_{\xi_1}=(\xi_1, \alpha_1)$ and $X_{\xi_2}=(\xi_2, \alpha_2)$, and let $x_{\xi_1},x_{\xi_2}$ be the corresponding dual elements. The Kirillov-Kostant symplectic form through dual element $(T, \beta)$ is
\begin{align}
    \omega(x_{\xi_1},x_{\xi_2}) = \left<(T,\beta), [X_{\xi_1},X_{\xi_2}]\right> = \oint dz  \left[T (\xi_1 \xi_2'-\xi_1' \xi_2) + \frac{\beta}{48\pi} (\xi_1 \xi_2''' - \xi_1''' \xi_2)\right]~.
\end{align}

Focusing now on the case of our non-smooth generalization of the Virasoro algebra, we can define the coadjoint orbit $\mathcal{O}_{T_*}$ through the unorthodox element $T_* = (P_0,c)$ of the dual space defined by the projection operator, Eq.~\eqref{eq:projectionoperator2}, together with its central extension $c$ in the full algebra. Again considering elements $x_{\xi_1},x_{\xi_2}$ in the dual space that correspond to algebra elements $X_{\xi_1},X_{\xi_2}$ through the pairing, and using Eq.~\eqref{eq:projectioninz}, this becomes
\be 
\omega(x_{\xi_1},x_{\xi_2}) = \left<T_*, [X_{\xi_1},X_{\xi_2}]\right> = P_0([X_{\xi_1},X_{\xi_2}]) + \frac{c}{48\pi} \oint dz \left[ (\xi_1 \xi_2''' - \xi_1''' \xi_2)\right]~.
\ee
This is precisely Eq.~\eqref{eq:berrycurvaturec} for the curvature. Thus, the modular Berry curvature for state-changing parallel transport is now related to the symplectic form on this orbit. 

What is the holographic bulk interpretation of such a non-standard orbit? We will argue that the corresponding geometry is related to the backreaction of a cosmic brane. 

%% file: gravity.tex
\section{Bulk phase space interpretation}\label{sec:bulk}
A Berry curvature for pure states constructed from Euclidean path integrals was shown to be equal to the integral of the bulk symplectic form over a Cauchy slice extending into the bulk in \cite{Belin:2018fxe,Belin:2018bpg} (see also~\cite{Marolf:2017kvq}). The notion of Uhlmann holonomy is one particular generalization of Berry phases to mixed states, and it was argued in \cite{Kirklin:2019ror} that its holographic dual is the integral of the bulk symplectic form over the entanglement wedge. However, the arguments for arriving at this result for Uhlmann holonomy are purely formal, and to the best of our knowledge this identification has not been worked out in an explicit example. The derivation also lacks a precise definition for the entanglement wedge symplectic form, which we will provide.

In this section, we will comment on a possible bulk interpretation of the modular Berry curvature for state-changing parallel transport. We will see that the result for the curvature that we obtained in the previous sections is closely related to an integral of a bulk symplectic form on a geometry with a conical singularity. See~\cite{Lewkowycz:2013nqa, Dong:2016fnf, Dong:2013qoa, Hung:2011nu, Krasnov:2001cu} for a related discussion of this geometry. 

\subsection{The conical singularity geometry} \label{sec:conicalsing}
We consider a Euclidean geometry obtained through the backreaction of a codimension-2 brane homologous to the boundary interval $A$. This leads to a family of Euclidean bulk solutions, which we denote by $\mathcal{M}_n$, where $n$ is a function of the tension of the brane~\cite{Dong:2016fnf}:
\be
\mathcal{T}_n=\frac{n-1}{4n G}~.
\ee
In the limit $n\to 1$, the cosmic brane becomes tensionless and settles on the location of the the usual RT surface associated to the entangling region, but for non-zero tension the brane backreacts on the geometry. The resulting geometries $\mathcal{M}_n$ are used in the context of the holographic computation of R\'enyi entropies $S_n$ in the boundary CFT, and we will argue that these are also relevant for a holographic interpretation of the modular Berry curvature.  

Let us first examine the boundary dual of the backreaction process. Inserting a cosmic brane which anchors the boundary at $z_1$ and $z_2$ corresponds to the insertion of twist fields $\mathcal{O}_{n}$ in the CFT at $z_1$ and $z_2$ \cite{Hung:2011nu}. The field $\mathcal{O}_{n}(z)$ is a (spinless) conformal primary of dimension~\cite{Calabrese:2004eu}
\be
\Delta_n=\frac{c}{12}\left(n-\frac{1}{n}\right)~.
\ee 
We use the fact that the cosmic brane can be computed as a correlation function of $\mathbb{Z}_n$ twist operators $\mathcal{O}_n, \mathcal{O}_{-n}$ in the boundary theory \cite{Hung:2011nu,Dong:2016fnf}.
\begin{figure}[t!]
\centerline{\includegraphics[scale=.3]{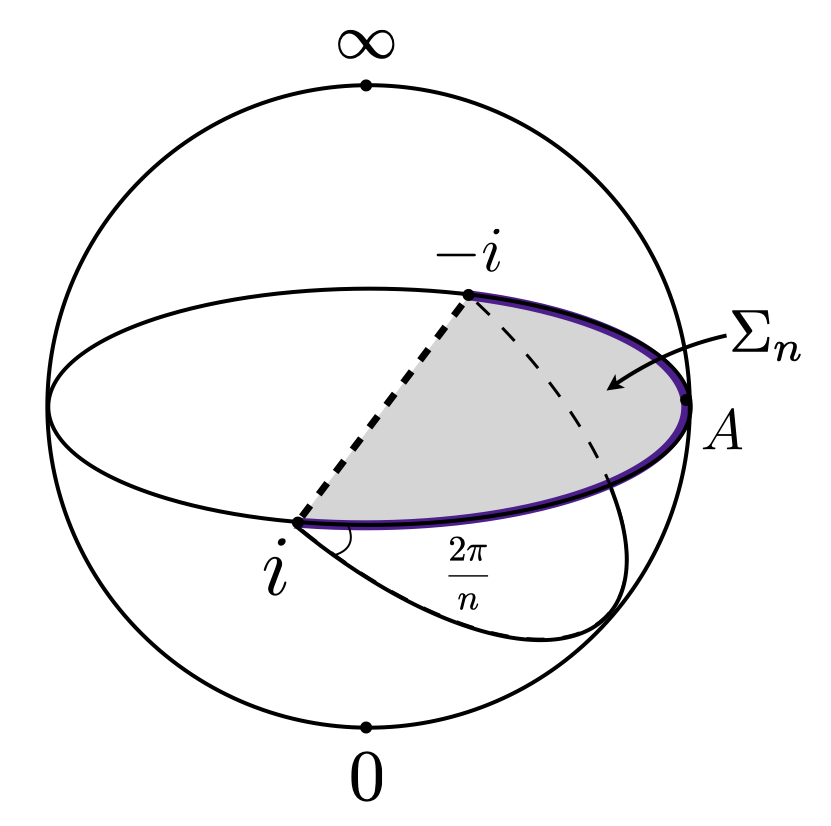}}
\caption{The conical singularity geometry $\mathcal{M}_n$ and entanglement wedge region $\Sigma_n$ corresponding to the boundary region $A$. The thick striped line corresponds to the cosmic brane extending from $-i$ to $i$. The backreaction process creates a conical singularity of opening angle $2\pi/n$.}
\label{fig:EntanglementWedge}
\end{figure}

Geometrically, we can think about the twist field as creating a conical singularity at the insertion point. Let us denote the two-dimensional geometry obtained from $\mathcal{O}_{n}(z_1), \mathcal{O}_{-n}(z_2)$  by $\mathcal{B}_n$. We are interested in the stress tensor profile on the boundary of the backreacted geometry, which by this reasoning is given by the  stress tensor on the plane in the background of two twist fields:  \begin{equation} \label{eq:twistfields}
\langle T(z)\rangle_{\mathcal{B}_n}= \frac{\langle T(z)\mathcal{O}_n(z_1)\mathcal{O}_{-n}(z_2)\rangle_{\mathbb{C}} }{\langle\mathcal{O}_n(z_1)\mathcal{O}_{-n}(z_2)\rangle_{\mathbb{C}} }~. 
\end{equation}
Using the general form of the three-point function in a CFT in terms of conformal dimensions, it now follows that $T(z)$ has poles of order two at $z_1$ and $z_2$ respectively.

To describe the geometry $\mathcal{M}_n$ explicitly, we consider the complex plane with coordinate $z$ which is flat everywhere except for two conical singularities at $z=z_1$ and $z=z_2$. The singular points are assumed to have a conical deficit of magnitude  
\begin{equation}
\Delta \varphi = 2\pi\left(1-\frac{1}{n}\right).
\end{equation}
We can use a uniformizing function $f(z)$ to map the $z$-plane with conical singularities to the smooth covering space, which we denote by $\widetilde{\mathcal{B}}_n$, which is a complex plane with coordinate $z'$ defined by
\begin{equation} \label{eq:conical}
z'=f(z)=\left(\frac{z-z_1}{z-z_2} \right)^{\frac{1}{n}}~.
\end{equation}
This maps $z_1 \to 0$ and $z_2 \to \infty$ so that the interval between $z_1$ and $z_2$ goes to the positive real axis $[0,\infty)$. The power of $\frac{1}{n}$ removes the conical singularity by gluing the $n$ sheets of the $z$-plane together, each represented by a wedge of opening angle $\frac{2\pi}{n}$. 

In terms of the coordinate $z'$ we extend $\widetilde{\mathcal{B}}_n$ into the bulk by introducing a `radial' coordinate $w'$ with metric of the form 
\be \label{eq:metrichyperbolic}
ds^2=\frac{dw'^2+dz'd\bar{z}'}{w'^2}~.
\ee
Here, we restrict the range of $z'$ by the identification $z'\sim e^{2\pi i/n}z'$, as this represents a fundamental domain $\widetilde{\mathcal{B}}_n/\mathbb{Z}_n$ in the covering space. The bulk coordinate approaches the boundary in the limit $w'\to 0$. The metric in Eq.~\eqref{eq:metrichyperbolic} is a wedge of three-dimensional hyperbolic space $\mathbb{H}^3$. We now use the following transformation:
\be 
w'= w \frac{1}{N}\sqrt{f'(z)\bar{f}'(\bar{z})}~,\hspace{10pt} z'=f(z)-w^2\frac{1}{N}\frac{f'(z)\bar{f}''(\bar{z})}{2\bar{f}'(\bar{z})}~,
\ee
where $f(z)$ is defined in Eq.~\eqref{eq:conical} and 
\be 
N=1+w^2\frac{f''(z)\bar{f}''(\bar{z})}{4f'(z)\bar{f}'(\bar{z})}~.
\ee
This transformation reduces to the conformal transformation in Eq.~\eqref{eq:conical} when we go to the boundary $w\to 0$. The metric in the new coordinates reads
\be\label{eq:FGmetric}
ds^2 =\frac{dw^2}{w^2}+\frac{1}{w^2}\left(dz-w^2\frac{6}{c}\bar{T}(\bar{z})d\bar{z}\right)\left(d\bar{z}-w^2\frac{6}{c}T(z) dz\right)~,
\ee
where 
\be \label{eq:stresstensor}
T(z)=\frac{c}{12}\{f(z),z\}=\frac{c}{24}\left(1-\frac{1}{n^2}\right)\frac{(z_1-z_2)^2}{(z-z_1)^2(z-z_2)^2}~,
\ee
with a similar expression holding for the anti-holomorphic component of the stress tensor $\bar{T}(\bar{z})$. The metric Eq.~\eqref{eq:FGmetric} falls into the class of Ba\~{n}ados geometries \cite{Banados:1998gg}, and $T(z)$ has the interpretation of the expectation value of the stress tensor in the boundary CFT on $\mathcal{B}_n$. Therefore, Eq.~\eqref{eq:stresstensor} agrees with the expression, Eq.~\eqref{eq:twistfields}, in terms of twist fields. The formula for $T(z)$ can also be seen more directly from the way the stress tensor in a CFT transforms under a conformal transformation. Starting from the vacuum stress tensor in the $z'$-coordinate, $T(z')=0$, and applying Eq.~\eqref{eq:conical}, the transformation picks up precisely the Schwarzian contribution in Eq. \eqref{eq:stresstensor}. 

We can also give a description for these geometries in the language of Chern-Simons (CS) theory. It is known that Euclidean AdS$_3$ can be described by two copies of a Chern-Simons theory with gauge connections $A, \bar{A}$ valued in $\mathfrak{sl}(2,\mathbb{C})$, and where the Chern-Simons coupling is related to Newton's constant by $k=(4G_3)^{-1}$~\cite{cmp/1104202513}. We can expand these connections (with complex coefficients) over $\mathfrak{sl}(2,\mathbb{R})$ generators $L_0,L_\pm$ satisfying $[L_0,L_\pm] = \mp L_\pm$, $[L_+,L_-] = 2 L_0$. In an explicit two-dimensional representation of the algebra, these are
\begin{equation} \label{Lgenerators}
  L_0= \frac{1}{2}
   \begin{pmatrix}
   1 & 0\\
   0 & -1
    \end{pmatrix}~,
    \quad L_+= 
   \begin{pmatrix}
   0 & 0\\
   -1 & 0
    \end{pmatrix}~,
     \quad L_-=
   \begin{pmatrix}
   0 & 1\\
   0 & 0
    \end{pmatrix}~.
\end{equation}
We can then describe the geometries, Eq.~\eqref{eq:FGmetric}, using the connections
\be \label{eq:CSconnections}
A=\frac{1}{2w}\begin{pmatrix} dw& -2\, dz \\ w^2\frac{12}{c}T(z)\, dz & -dw \end{pmatrix}, \hspace{20pt} \bar{A}=-\frac{1}{2w}\begin{pmatrix} dw & w^2\frac{12}{c}\bar{T}(\bar{z})\,d\bar{z} \\ -2\,d\bar{z} & -dw \end{pmatrix}~.
\ee
Each metric in this family of solutions corresponds to a choice of gauge connections, Eq.~\eqref{eq:CSconnections}, with the same $T(z), \bar{T}(\bar z)$ through the relation $ds^2 =\frac{1}{2}\mbox{tr}((A-\bar{A})^2)$.

It will be useful to extract the radial dependence in Eq.~\eqref{eq:CSconnections} by using a suitable gauge transformation 

\be A = b a b^{-1} + b db^{-1}~, \indent \bar{A} = b^{-1} \bar{a} b + b^{-1}db~, \ee
with gauge parameters
\be \label{eq:gaugefieldbdy}
a=\begin{pmatrix} 0 & -dz \\ \frac{6}{c}T(z)\,dz & 0 \end{pmatrix}~, \hspace{10pt} \bar{a}=\begin{pmatrix} 0 & -\frac{6}{c}\bar{T}(\bar{z})\,d\bar{z} \\ d\bar{z} & 0 \end{pmatrix}~, \hspace{10pt}b(w)=\begin{pmatrix} \frac{1}{\sqrt{w}} & 0 \\ 0 & \sqrt{w} \end{pmatrix}~.
\ee

\subsection{Symplectic form}\label{sec:sympform}

We now turn our attention to the bulk symplectic form. It is useful to work in the Chern-Simons formulation of three-dimensional gravity. For a similar discussion of the symplectic structure of $3$d gravity in this setting, especially as pertains to the connection to coadjoint orbits, see~\cite{Cotler:2018zff, Barnich:2017jgw, Coussaert:1995zp, Kraus:2021cwf}.

The CS action with CS coupling $k$ and gauge connection $A$ is given by
\be S_{\rm CS} = \int \mathcal{L}_{\rm CS} = \frac{k}{4\pi}\int \mbox{tr}\left(A \wedge dA + \frac{2}{3} A\wedge A \wedge A\right)~.\ee
We would like to evaluate the symplectic form. Taking the variation of the action for a single copy gives
\be \delta \mathcal{L}_{\rm CS} =  \frac{k}{2\pi}\mbox{tr}\left(\delta A \wedge F\right) + d\Theta ~\ee
in terms of field strength $F = dA + A\wedge A$, and where $\Theta = \frac{k}{4\pi}\mathrm{tr}(A\wedge \delta A)$. The symplectic form for CS theory on some spatial region $\Sigma$ is then given by
\be
\omega= \int_\Sigma \delta \Theta = \frac{k}{4\pi}\int_{\Sigma} \mathrm{tr}(\delta_1 A \wedge \delta_2 A)~.
\ee
In the following, we will assume that $\Sigma$ is topologically a disk, i.e., it has a single boundary but no singularities in the interior. The symplectic form is a two-form on the space of classical solutions satisfying $F=0$. Because we are working with a disk which admits no nontrivial cycles, a variation $\delta A$ which leaves this condition invariant is of the form 
\be  \label{eq:deltaA}
\delta A =d_A\zeta\equiv d\zeta+[A,\zeta]
\ee
for some gauge transformation $\zeta$, as follows from $\delta F = d_A \delta A = d_A^2 \zeta = 0$. 

We now consider the symplectic form for such a transformation. Using the identity
\be 
\mathrm{tr}\left([A,\zeta] \wedge \delta A\right)=-\mathrm{tr}\left(\zeta\wedge [A,\delta A]\right)~
\ee 
and integrating by parts we obtain
\begin{align}
\omega&=\frac{k}{4\pi}\int_{\Sigma} \mathrm{tr}(d_A\zeta \wedge \delta A)=\frac{k}{4\pi}\oint_{\partial \Sigma} \mathrm{tr}(\zeta \wedge \delta A)-\frac{k}{4\pi}\int_{\Sigma} \mathrm{tr}(\zeta \wedge d_A\delta A) \nonumber \\ \label{eq:sympform}
&=\frac{k}{4\pi}\oint_{\partial \Sigma} \mathrm{tr}(\zeta \wedge \delta A)~.
\end{align}
From Eq.~\eqref{eq:sympform} we see that the symplectic form $\omega$ is localized at the boundary of $\Sigma$.

Suppose that $\partial \Sigma$ lies in the asymptotic boundary of the geometry, in the $w=0$ plane, and that we have gauged away the radial dependence. Using the explicit form of the connections, Eqs.~\eqref{eq:CSconnections} and~\eqref{eq:gaugefieldbdy}, we can evaluate the symplectic form in Eq.~\eqref{eq:sympform}. We see that the field variation can be expressed in terms of the stress tensor as 
\be \label{eq:deltaA2}
\delta A =\frac{6}{c}\begin{pmatrix} 0 & 0 \\ \delta T & 0 \end{pmatrix} dz~.
\ee
It is also possible to solve Eq.~\eqref{eq:deltaA} for $\delta T$. Decomposing $\zeta$ over the $\mathfrak{sl}(2,\mathbb{R})$ generators as $\zeta=\zeta_{-}L_{-1}+\zeta_0L_0+\zeta_+L_{1}$
and using the form of the gauge field in Eq.~\eqref{eq:gaugefieldbdy}, one can compute $d_A\zeta$.  Matching with Eq.~\eqref{eq:deltaA2} gives a solution of the form
\be
\delta T = \frac{c}{12}\xi''' + 2 T \xi' + \partial T \xi~, \label{eq:stresstransform}
\ee 
where we have written $\xi\equiv-\zeta_{-}$ for the component of the gauge transformation associated to the $L_{-1}$ generator. This is the usual stress tensor transformation law. From the form of the gauge transformation and the variation $\delta A$ in Eq.~\eqref{eq:deltaA}, and using the Brown-Henneaux relation $(4G_3)^{-1} = c/6$ combined with the gravitational value for the CS coupling, we find that 
\be
\omega=\frac{1}{4\pi}\oint_{\partial \Sigma} dz\, \xi \wedge \delta T~.
\ee
Using Eq.~\eqref{eq:stresstransform} the symplectic form becomes
\be
\omega=\frac{1}{4\pi}\oint_{\partial \Sigma} dz \left( \frac{c}{12}\xi \wedge \xi''' +2 T\, \xi \wedge \xi' \right).
\ee
Plugging in two diffeomorphisms $\xi_1$ and $\xi_2$, the final result for the symplectic form reads: 
\be \label{eq:sympformfinal}
\omega= \frac{1}{2\pi}\oint_{\partial \Sigma} dz\, \left( T \left(\xi_1\xi_2'-\xi_2\xi_1'\right)+\frac{c}{24}\left( \xi_1\xi_2'''-\xi_2\xi_1'''\right) \right)~.
\ee
When the stress tensor $T(z)=T$ is a constant, Eq.~\eqref{eq:sympformfinal} is reminiscent of the Kirillov-Kostant symplectic form on the coadjoint orbit $\rm{Diff}(S^1)/U(1)$ (or $\rm{Diff}(S^1)/SL(2,\mathbb{R})$ for the vacuum stress tensor) of the Virasoro group $\widehat{\rm{Diff}(S^1)}$ with central charge $c$. However to match onto the Berry curvature, Eq.~\eqref{eq:berrycurvature}, with the zero mode projection Eq.~\eqref{eq:projectioninz}, we must consider a non-constant vacuum stress tensor. In fact the the stress tensor profile that reproduces the correct projection is of the form Eq.~\eqref{eq:stresstensor}. In other words, the zero mode projection for the parallel transport process is implemented by integrating against the stress-tensor expectation value in the presence of two twist fields. We will now argue more precisely that in order to match the modular Berry curvature we need to consider a non-standard orbit corresponding to the conical singularity geometry described in Section \ref{sec:conicalsing}. 

\subsection{Contour prescription}

Let us return to the Euclidean geometry $\mathcal{M}_n$, which is obtained from the backreaction of a cosmic brane with tension $\mathcal{T}_n$. We showed that the stress tensor profile at the boundary is given by Eq.~\eqref{eq:stresstensor}. Let us now restrict to transformations which leave the interval at the boundary fixed. This corresponds to Dirichlet boundary conditions $\delta A=0$ at the cosmic brane. 

We consider the symplectic form 
\begin{equation}
\omega_{n}=\frac{k}{4\pi}\int_{\Sigma_n}\mathrm{tr}(\delta_1 A \wedge \delta_2 A)~,
\end{equation}
supported on some region $\Sigma_{n}$ which corresponds to the entanglement wedge in the geometry $\mathcal{M}_n$, see Figure \ref{fig:EntanglementWedge}. The subscript in the symplectic form indicates that it depends on $n$. The entanglement wedge has two boundary components: 
\begin{equation}
\partial \Sigma_{n} = \gamma_n \cup \mathrm{Brane}_n~,
\end{equation} 
where $\gamma_n$ is the entangling region at the asymptotic boundary extending between $z_1$ and $z_2$ and $\mathrm{Brane}_n$ is the cosmic brane anchored at those points. In Section~\ref{sec:sympform}, we have seen that the bulk symplectic form localizes to the boundary of $\Sigma_n$ (using that the region is topologically trivial), because $\mathrm{tr}(\delta_1 A \wedge \delta_2 A)=d\eta$ is an exact form with $\eta=\mathrm{tr}(\xi \wedge \delta A)$. The expression for $\omega_n$ therefore reduces to a boundary term of the form 
\begin{equation}
\omega_n =\frac{k}{4\pi}\left[\int_{\gamma_n}\eta+\int_{\mathrm{Brane}_n}\eta\right]~.
\end{equation}
The contribution at the cosmic brane vanishes due to the boundary conditions we put on the field variations there, i.e., $\delta A=0$ at $\mathrm{Brane}_n$. We are therefore left with the integral over the entangling region $\gamma_n$ at the asymptotic boundary. There, $\eta$ takes the form 
\begin{equation}
k \, \eta = \xi \wedge \delta T =\frac{c}{12}\left(\xi_1\xi_2'''-\xi_2\xi_1'''\right)+2\, T[\xi_1,\xi_2]~,
\end{equation}
in terms of the boundary stress tensor profile $T$ of the geometry $\mathcal{M}_n$. Plugging in Eq.~\eqref{eq:stresstensor} with $z_1=i$ and $z_2=-i$, we find that 
\be \label{eq:omegan}
\omega_n =\frac{c}{12\pi}\left(1-\frac{1}{n^2} \right)\int_{\gamma_n}\frac{[\xi_1,\xi_2]}{(z^2+1)^2}\, dz+\frac{c}{48\pi}\int_{\gamma_n}\left(\xi_1\xi _2'''-\xi_2\xi_1'''\right)dz~.
\ee
Note that the integrand is singular at the endpoints of the integration region $\gamma_n$. Therefore, we should implement some kind of regularization procedure for the integral to avoid the twist field insertion points. A standard choice would be the \emph{principal value} prescription, where we excise a small ball of size $\epsilon$ around each of the singularities located at the endpoints of $\gamma_n$. After computing the integral, we take $\epsilon \to 0$. The resulting expression for $\omega_n$ is UV divergent ($\omega_n \sim \log \epsilon$). 

In the limit $n\to 1$ the first term in Eq.~\eqref{eq:omegan} vanishes. This is expected, since as the cosmic branes becomes tensionless the geometry reduces to pure $\mathrm{AdS}_3$, for which the bulk symplectic form is identically zero (up to the central charge term). To extract a non-zero answer from $\omega_n$, we first take a derivative with respect to $n$ and define
\be \label{eq:bulksymp}
\omega \equiv \lim_{n\to 1}\frac{\partial}{\partial n} \frac{\omega_n}{k}~.
\ee
This corresponds to studying the first order correction of the backreaction process. The appearance of the operator $\lim_{n\to 1}\partial_n$ is not unfamiliar in the context of computing entanglement entropy using Euclidean solutions with conical singularties of the form $\mathcal{M}_n$\footnote{In fact, the entanglement entropy $S$ associated to the subregion $A$ can be computed by the formula $S=-\lim_{n\to 1}\partial_n\log Z_n$, where $\log Z_n\sim -I\left[\mathcal{M}_n\right]$ is the classical action evaluated on the conical singularity geometry $\mathcal{M}_n$.}. Eq.~\eqref{eq:bulksymp} is our proposal for the bulk symplectic form associated to the entanglement wedge, and we will now show that it matches the modular Berry curvature.

To make the connection with the boundary computation, we rewrite the integral over the entangling region in terms of the variable $u$ defined in Eq.~\eqref{eq:coordtrans}. Notice that the unit semicircle $-\pi/2\leq \arg(z)\leq\pi/2$ is mapped to the line $u\in [-\infty,\infty]$, since $z=1$ goes to $u=0$. In particular, the points $u=\pm \Lambda$ correspond to
\be \label{eq:endpoints}
z=\frac{1+ie^{\pm \Lambda}}{e^{\pm \Lambda}+i} \sim e^{\pm i\left(\frac{\pi}{2} - \epsilon\right)}~,
\ee
if we identify $\Lambda$ with the UV regulator by $\Lambda = -\log\frac{\epsilon}{2}$, in the limit $\Lambda\rightarrow \infty,\epsilon\rightarrow 0$. In the limit $\Lambda \to \infty $, the endpoints go to $z\to \pm i$ along the unit circle, so Eq.~\eqref{eq:endpoints} is precisely the principal value prescription for $\gamma_n$. 

Moreover, under the transformation in  Eq.~(\ref{eq:coordtrans}) the integration measure changes as Eq.~\eqref{eq:uzcoordtransf}. Therefore, we can represent the integral over the entangling region $\gamma_n$ in terms of the $u$-variable as
\begin{equation}
\frac{1}{\pi}\int_{-i}^{i}\frac{\xi(z)}{(1+z^2)^2}\,dz=\lim_{\Lambda\to \infty}\frac{i}{2\pi}\int_{-\Lambda}^{\Lambda}\xi(u)\, du~,
\end{equation}
which is precisely the projection operator $P_0(X_{\xi})$ in Eq.~\eqref{eq:projectionoperator}. Thus, we can rewrite the symplectic form $\omega_n$ as
\be
\omega_n =\frac{c}{12}\left(1-\frac{1}{n^2} \right)P_0(\left[X_{\xi_1},X_{\xi_2}\right])+\frac{c}{48\pi} \int_{-i}^i \left(\xi_1\xi _2'''-\xi_2\xi_1'''\right)dz~.
\ee
Taking the derivative with respect to $n$ and setting $n\to 1$ according to Eq.~\eqref{eq:bulksymp} gives the final result:
\be
\omega = P_0(\left[X_{\xi_1},X_{\xi_2}\right])~,
\ee
which agrees with the curvature $F$ in Eq.~(\ref{eq:berrycurvature}). Notice that the information about the central zero mode discussed in Section \ref{sec:extension} is also contained in $\omega_n$: it simply corresponds to taking $\lim_{n\to 1} \omega_n$ directly.

%% file: discussion.tex
\section{Discussion}

We have considered the case of boundary parallel transport of a fixed interval under a change in global state, which is in contrast to the situation considered in~\cite{Czech:2019vih} where the state is held fixed while the interval location is varied. However, a general parallel transport process will change both the state and the location of the interval. In such a situation, the curvature will contain cross-terms between the $X_\lambda$'s of Eq.~\eqref{eq:projectionoperator} and the $V_\mu$'s of Section~\ref{sec:IntervalTransport}. Both are eigenoperators of the adjoint action of the modular Hamiltonian, $[H_{\rm mod}, X_\lambda] = \lambda X_\lambda$ and $[H_{\rm mod}, V_\mu] = i \mu V_\mu$, but notice that the eigenvalue of the $X_\lambda$'s is real while that of $V_\mu$ is purely imaginary. By the Jacobi identity, the commutator $[X_\lambda, V_\mu]$ will have an eigenvalue that is the sum of the two, thus it has both a real and imaginary part. This is never zero, which means $[X_\lambda, V_\mu]$ does not have a zero mode. The curvature, Eq.~\eqref{eq:berrycurvature}, is given by the projection onto this zero mode, which means that computed in these directions that mix changes of state and interval location, it must vanish. Thus, it appears to be sufficient to consider state and interval location-based transport separately. 

In the bulk, we have demonstrated an abstract connection between state-changing parallel transport of boundary intervals and a certain family of Euclidean bulk solutions. The holographic dual of the modular Berry curvature was argued to be an entanglement wedge symplectic form on this geometry. This is similar in spirit to the results of \cite{Belin:2018bpg,Belin:2018fxe}, but in the case of mixed states. However, a direct phase space interpretation of this symplectic form in Lorentzian signature is not so obvious. Associating a phase space, i.e., a solution space of a proper initial value problem, to an entanglement wedge involves some subtleties, e.g., the possibility of edge modes \cite{Donnelly:2016auv,Speranza:2017gxd,Donnelly:2020xgu}  and boundary ambiguities at the RT surface that must be fixed by a suitable choice of boundary conditions. Possibly, one could exploit the relation to the hyperbolic black hole and identify the relevant phase space with the one associated to the (outside of the) black hole. This would lead to geometric setup for which the Lorentzian continuation is more well-behaved. In particular, this approach requires a further study of the choice of boundary conditions that are natural to put at the horizon. 

\begin{figure}[t!]
\centerline{\includegraphics[scale=.5]{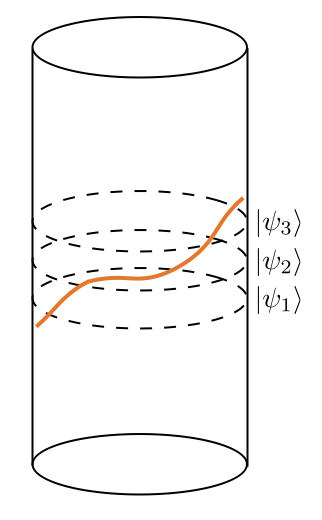}}
\caption{An example of a time-dependent geometry limiting to different boundary states $\ket{\psi_i}$ at each time. Could the Berry phase associated to state-dependent parallel transport compute the length of a curve (such as the thick orange curve) in such a geometry?}
\label{fig:gluing}
\end{figure}

It would also be interesting to explore a bulk description within a single Lorentzian geometry. For instance, one could imagine constructing a time-dependent geometry by gluing together certain slowly varying time-independent geometries that are each dual to different boundary states. Since this will not in general give an on-shell solution, one could try to turn on suitable sources on the boundary as a function of time, in such a way that time evolution
under the modified Hamiltonian (with sources) provides precisely the sequence of states under consideration. In such a situation, one could look for a corresponding on-shell bulk solution with modified asymptotics. It would be interesting to explore whether the Berry phase associated to state-changing parallel transport computes a length within a time-dependent geometry (see Figure~\ref{fig:gluing}).

Additionally, it would be interesting to explore further the connections to Uhlmann holonomy described in~\cite{Kirklin:2019ror}. This is a version of parallel transport constructed from purification of density matrices subject to certain maximization conditions on transition probabilities. Through appropriate insertion of stress tensors at the boundary, this is claimed in \cite{
Faulkner:2015csl,Faulkner:2016mzt, Lewkowycz:2018sgn,Banerjee:2011mg} to describe the shape-changing transport problem considered in Section~\ref{sec:IntervalTransport}. In this setting, the Berry curvature associated to a parallel transport process that changes the state was argued to be dual to the symplectic form of the entanglement wedge. While similar in spirit to much of this work, it would be interesting to further study the relation to our work in the context of key differences, such as the need for diagonalizing the adjoint action and the use of non-smooth vector fields. 

The problem we study also has relevance for thermalization in 2d CFT. For example, the Krylov complexity contains information about operator growth in quantum chaotic systems. Roughly speaking, this is given by counting the operators that result under nested commutators with respect to a `Hamiltonian' of the system. In~\cite{Caputa:2021ori}, the Krylov complexity was studied for the case where this Hamiltonian takes the form of Eq.~\eqref{eq:HmodVir}, using an oscillator representation of the Virasoro algebra. This is similar to the modular Berry transport process we have considered, with the exception again of the use of non-smooth vector fields.

In studying operator-based parallel transport, we uncovered some subtleties regarding the diagonalization of the adjoint action for arbitrary Virasoro generators (an explanation of these issues was given in Appendix~\ref{sec:NonDiagonalization}). For this reason we considered a set of certain non-smooth vector fields on the circle, Eq.~\eqref{eq:eigenfunction}, which explicitly diagonalize the adjoint action so that the curvature results of Appendix~\ref{app:general} may be applied. It would be interesting to further study this issue. For instance, we found that the adjoint action could not be diagonalized over the usual Virasoro algebra, defined as the set of smooth vector fields on the circle.\footnote{This is actually not uncommon in the case of infinite-dimensional vector spaces. For example, when one tries to diagonalize the derivative operator on the space of polynomial functions one naturally finds exponential functions, which are not part of the original space. The non-analyticities we found should be regarded in the same way.} Instead, we saw that the set of generators not expressible as $[H_{\rm mod}, X]$ was dimension three, larger than the dimension of the kernel (which is in this case one-dimensional and generated by $H_{\rm mod}$). Furthermore, there was an ambiguity in the non-zero mode piece. One could ask whether it is possible to consider parallel transport generated by elements of the usual Virasoro algebra, and perhaps resolve the ambiguities in the decomposition by taking a suitable choice of norm. Along these lines, one could consider only Virasoro algebra elements that are contained within physical correlators. It would be interesting to apply techniques from algebraic quantum field theory to see if this eliminates some of the ambiguities we have encountered. 

To properly diagonalize the adjoint action we were led to consider vector fields on the circle that are non-differentiable on the endpoints of the interval. These form a continuous version of the Virasoro algebra. Our Berry curvature can be understood formally as the Kirillov-Kostant symplectic form on an orbit associated to this algebra. It would be interesting to conduct a more rigorous study of this algebra and its central extension. It is also worth noting that we considered a dual space of distributions on the circle, which is larger than the set of smooth quadratic differentials considered in the classification of~\cite{Witten:1987ty}. For this reason, the orbits we consider differ considerably from known Virasoro orbits since the associated representative, Eq.~\eqref{eq:projectioninz2}, is not a quadratic form on the circle. To our knowledge, such orbits have not been studied before in the literature. We have identified at least one physical implication of such unconventional orbits, and thus it would be interesting to revisit the classification of Virasoro orbits using more general duals.

%% file: appendix.tex
\section{Kinematic space example}\label{app:kinematic}

We will now describe a version of the state-based parallel transport summarized in Section~\ref{sec:states}, which reproduces some of the results from kinematic space for CFT$_2$ on a time-slice. As we saw in Section~\ref{sec:IntervalTransport}, the parallel transport process for kinematic space could also be derived in the operator-based transport language. In this way of formulating the problem, the geometrical description of kinematic space in terms of coadjoint orbits~\cite{Penna:2018xqq} is more readily transparent.

We will start by setting up some geometry that is relevant for this problem. Consider the group $SL(2,\mathbb{R})$. Its Lie algebra $\mathfrak{sl}(2,\mathbb{R})$ consists of generators $t_\mu$, $\mu=0,1,2$ satisfying the commutation relations $[t_\mu,t_\nu]=\epsilon_{\mu \nu}\,^\rho t_\rho$, where the indices are raised by a metric $\eta_{ab}$ with signature $(-,+,+)$. We will make use of an explicit finite-dimensional representation by $2\times 2$ matrices given by
\begin{equation} \label{generators}
  t_0= \frac{1}{2}
   \begin{pmatrix}
   0 & 1\\
   -1 & 0
    \end{pmatrix}~,
    \quad t_1= \frac{1}{2}
   \begin{pmatrix}
   0 & 1\\
   1 & 0
    \end{pmatrix}~,
     \quad t_2= \frac{1}{2}
   \begin{pmatrix}
   1 & 0\\
   0 & -1
    \end{pmatrix}~.
\end{equation}
This basis will be most convenient for the calculation of the Berry curvature. It can be easily expressed in terms of the basis used in Section~\ref{sec:bulk} as $t_0=\frac{1}{2}(L_-+L_+),t_1=\frac{1}{2}(L_--L_+),t_2=L_0$.

Now consider embedding coordinates $(X^0,X^1,X^2)$ describing $3$-dimensional Minkowski spacetime
with metric
\begin{equation}
    ds^2=-(dX^0)^2+(dX^1)^2+(dX^2)^2~. \label{eq:MinkowskiMetric}
\end{equation}
Recall that $SL(2,\mathbb{R})/\mathbb{Z}_2\cong SO(2,1)$. A convenient parametrization for the algebra $\mathfrak{sl}(2,\mathbb{R})$ is given through the isomorphism to Mink$_3$:
\begin{equation}
    \frac{1}{2}\begin{pmatrix}
    X^2& X^1+X^0\\
    X^1-X^0& -X^2
    \end{pmatrix} \leftrightarrow (X^0,X^1,X^2)~.\label{eq:MinkowskiMap}
\end{equation}
The reason to express $\mathfrak{sl}(2,\mathbb{R})$ in this way is that the coadjoint orbits of the Lie group can be realized geometrically in Minkowski space. Any element of  $\mathfrak{sl}(2,\mathbb{R})$ lies in one of three conjugacy classes (up to an overall factor $\pm 1)$. These can be classified by the value of $\epsilon \equiv |\mbox{tr}(g)|/2$ where $g\in SL(2,\mathbb{R})$: $\epsilon<1$ is elliptic, $\epsilon=1$ is parabolic and $\epsilon>1$ is hyperbolic.

\begin{figure}[t!]
\centerline{\includegraphics[scale=.35]{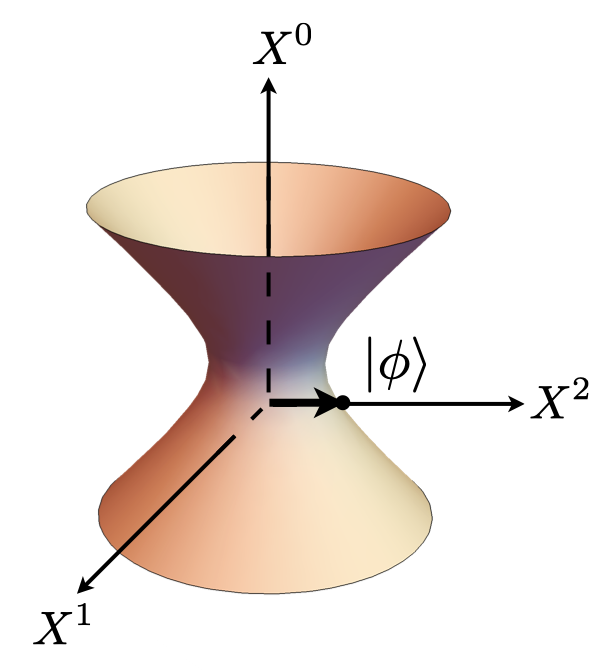}}
\caption{The dS$_2$ hyperboloid describing kinematic space, which is a coadjoint orbit of $SO(2,1)$. The arrow points to a special point that corresponds to the coherent state $\ket{\phi}$.}
\label{fig:dSplot}
\end{figure}

We will assume that our representative is in the diagonal class 
\be
\Lambda=\rm{diag}(\lambda,-\lambda)/2~\label{eq:representative}
\ee 
with $\lambda \in \mathbb{R}$. Since $|\mbox{tr}(e^\Lambda)|/2>1$ for all $\lambda$, this is a hyperbolic element. Other choices lead to different orbits. 

Consider a general group element
\be \label{eq:elementsl2r}
g=\begin{pmatrix} a & b\\ c& d \end{pmatrix} \in SL(2,\mathbb{R})~,
\ee
with $a,b,c,d\in \mathbb{R}$ and $ad-bc=1$. The coadjoint orbit is generated by the adjoint action of $\Lambda$ with arbitrary $g$,
\begin{equation} \label{eq:conjugation}
    g \cdot \Lambda \cdot g^{-1} = \begin{pmatrix}
    \frac{\lambda}{2}(bc+ad)&-\lambda ab\\
     \lambda cd & -\frac{\lambda}{2}(bc+ad)
    \end{pmatrix}~.
\end{equation}
The determinant is constant along the orbit, $\mbox{det}(g \cdot \Lambda \cdot g^{-1}) = -\lambda^2/4~.$ Applying the map to Minkowski space, Eq.~\eqref{eq:MinkowskiMap}, this results in the condition
\begin{align}
    -(X^0)^2+(X^1)^2+(X^2)^2&=\lambda^2~.\label{eq:EmbeddingConstraint}
\end{align}
This is the defining equation of a single-sheeted hyperboloid with radius $\lambda$.
Take the embedding coordinates
\begin{align}
X^0 &= \lambda \cot{t}~,\nonumber\\
X^1 &= \lambda \csc{t} \cos{\theta}~,\nonumber\\
X^2 &= \lambda \csc{t} \sin{\theta}~.\label{eq:Embedding}
\end{align}
These satisfy Eq.~\eqref{eq:EmbeddingConstraint} and from Eq.~\eqref{eq:MinkowskiMetric} result in the induced metric
\be ds^2 = \lambda^2 \csc^2{t} (-dt^2 + d\theta^2)~.\label{eq:dS2metric}\ee
This is just the metric on $\mbox{dS}_2\simeq SO(1,2)/SO(1,1)$. We saw that this describes the coadjoint orbit passing through the representative, Eq.~\eqref{eq:representative}.

The coadjoint orbit can be thought of as a fiber bundle whose base space is $SO(1,2)/SO(1,1)$ and its fiber is $SO(1,1)$. We want to consider an appropriate section of the fiber bundle. The discussion below follows closely \cite{Oblak:2017ect}. Using the embedding coordinate Eq.~\eqref{eq:Embedding} and the map Eq.~\eqref{eq:MinkowskiMap}, we obtain the constraints
\begin{align} 
    2\mathrm{tr}(g\, \Lambda\, g^{-1}\, t_0)&=-X_0 = -\lambda\cot{t}~,\nonumber \\
   2\mathrm{tr}(g\, \Lambda\, g^{-1}\, t_1)&= X_1 = \lambda\cos{\theta}\csc{t}~,\nonumber\\
    2\mathrm{tr}(g\, \Lambda\, g^{-1}\, t_2)&= X_2 = \lambda\sin{\theta}\csc{t}~. \label{eq:systemequations}
\end{align}
Solving this system of equations, Eq.~\eqref{eq:systemequations}, we obtain
\begin{equation}
   b=-\frac{\cot{t}+\cos{\theta}\csc{t}}{2a}~, \quad c=\frac{a(1-\sin{\theta}\csc{t})}{\cot{t}+\cos{\theta}\csc{t}}~, \quad d= \frac{1 +\sin{\theta}\csc{t}}{2a}~.
\end{equation}
We have the freedom to impose $a=1$, in which case the expressions somewhat simplify. Applying this back to Eq.~\eqref{eq:elementsl2r}, we obtain a section $g: \mbox{dS}_2 \rightarrow SL(2,\mathbb{R})$ for the bundle given by
\begin{equation} \label{section}
     g= \begin{pmatrix}
     1 & -\frac{1}{2}(\cos{t}+\cos{\theta})\csc{t}\\
     \tan{(\frac{t-\theta}{2})}& \frac{1}{2}(1+\csc{t}\sin{\theta})
     \end{pmatrix}~.
\end{equation}
Notice that $g$ reduces to the identity for $t=\theta=\pi/2$ which corresponds to the point of intersection of the hyperboloid with the axis labeled by the $t_2$ generator.

Now we will apply some of these tools to the problem of state-based parallel transport for the group $SL(2,\mathbb{R})$, with the aim of describing kinematic space. Recall that to define a state-based Berry phase it is necessary to choose a suitable `Hamiltonian' with an eigenstate $\ket{\phi}$ that serves as the base state for the parallel transport process. The `Hamiltonian' is one which generates a specified subgroup of $SL(2,\mathbb{R})$, which we interpret as a flow in time. The state is acted on by group elements in a unitary representation, which we denote by $\mathcal{D}(g)$, $\mathcal{D}(u)$ for $g\in SL(2,\mathbb{R}), u\in \mathfrak{sl}(2,\mathbb{R})$. In the coadjoint orbit language, eigenstates of subalgebras of the symmetry algebra are known as \emph{coherent states}. Specifically, we will choose our Hamiltonian to be $t_2$, which generates an $\mathfrak{so}(1,1)$ subalgebra. This exponentiates to the hyperbolic group element
\begin{equation}\label{expgen}
   \mathcal{J}= e^{\eta t_2/2}~
\end{equation}
with $\eta\in \mathbb{R}$. Taking $X\rightarrow \mathcal{J} X \mathcal{J}^{-1}$ using the isomorphism, Eq.~\eqref{eq:MinkowskiMap}, we see the adjoint action with respect to $\mathcal{J}$ acts geometrically as
\begin{align}
    X^0&\rightarrow X^0 \cosh{(\eta/2)} + X^1 \sinh{(\eta/2)}~,\\
    X^1&\rightarrow X^0 \sinh{(\eta/2)} + X^1 \cosh{(\eta/2)}~,\\
    X^2&\rightarrow X^2~,
\end{align}
in other words, it acts as a boost with rapidity $-\eta/2$ in the $X^0-X^1$ direction in embedding space.

We define our coherent state through the condition that the boost leaves it invariant up to a phase,
\begin{equation}\label{eq:coherenteigen}
    \mathcal{D}(\mathcal{J})\ket{\phi}=e^{i \eta \zeta}\ket{\phi}~, \indent \mathcal{D}(t_2)\ket{\phi}= 2\zeta\ket{\phi}~,
\end{equation}
with $\zeta\in \mathbb{R}$ since $\mathcal{D}(\mathcal{J})$ is assumed to be unitary and $\mathcal{D}(t_2)$ Hermitian in the representation. By a theorem of Perelomov~\cite{Perelomov:1986tf} (see also~\cite{Yaffe:1981vf}), coherent states are in one-to-one correspondence with points on an orbit. Our state $\ket{\phi}$ corresponds to the point $(0,0,1)$ on the dS$_2$ hyperboloid that is left fixed by the action of the boost (see Figure~\ref{fig:dSplot}). It is geometrically simple to see that the action of the other generators $t_0, t_1$ do not leave this point invariant, which corresponds to the statement that $\ket{\phi}$ is not also an eigenstate of these generators.

Recall that the Maurer-Cartan form is given by 
\be
\Theta=g^{-1}dg~.
\ee 
The Berry phase is
\begin{equation} \label{eq:thetaformula}
     \theta(\gamma)=\oint_\gamma{A}~, \quad A=i\bra{\phi}\mathcal{D}(\Theta)\ket{\phi}~.
\end{equation}
We now use Eq.~\eqref{section} to evaluate the pullback of the Maurer-Cartan form from $SL(2,\mathbb{R})$ to dS$_2$. Taking the expectation value of the generators in the state $\ket{\phi}$, then applying the commutation relations, the eigenvalue condition Eq.~\eqref{eq:coherenteigen} and using $\zeta\in \mathbb{R}$, we see that only $t_2$ has a nonvanishing expectation value in $\ket{\phi}$. Thus, only this part contributes to the Berry phase. We find
\begin{equation} \label{eq:integrand}
     A=i\bra{\phi}\mathcal{D}(\Theta)\ket{\phi}=i\zeta\csc{t}\cos{\left(\frac{t+\theta}{2}\right)}\sec{\left(\frac{t-\theta}{2}\right)}(dt-d\theta)~.
\end{equation}
From this we can define the Berry curvature 
\begin{equation}\label{Berrycurv}
F=dA=\frac{i\zeta}{\sin^2{t}} dt\wedge d\theta~.
\end{equation}
Using Stokes' theorem one can write the integral of the Berry connection in Eq.~\eqref{eq:integrand} as
\begin{equation}\label{Berryphase}
 \theta(\gamma)=i\zeta \int_B{\frac{1}{\sin^2{t}}dt\wedge d\theta}~,
\end{equation}
where $B$ is any two-dimensional region with boundary $\partial B =\gamma$. 

For a CFT$_2$ restricted to a time-slice, kinematic space consists of the space of intervals on this time-slice. Given a causal ordering based on containment of intervals, this is just a dS$_2$ spacetime, Eq.~\eqref{eq:dS2metric}, with a time coordinate set by the interval radius, $(\theta_R-\theta_L)/2$~\cite{Czech:2015qta}. The curvature, Eq.~\eqref{Berrycurv}, is a volume form on kinematic space. Recalling the relation between time and interval size, it matches the kinematic space curvature, Eq.~\eqref{eq:curvature3}, derived from the operator-based method in Section~\ref{sec:IntervalTransport} (note that an exact matching of the normalization is unimportant, as the overall normalization for the modular Berry phase will be at any rate affected by the choice of normalization for the modular Hamiltonian). The Berry phase, Eq.~\eqref{Berryphase}, computes the volume of region $B$ within this dS$_2$ spacetime. It also precisely reproduces the Berry phase for kinematic space derived by other means in~\cite{Czech:2019vih,Czech:2017zfq}.

\section{General formulation}\label{app:general}

We will derive a general formula for the curvature assuming that there is a unique way of separating out the zero mode. As we discuss in the next appendix, this is not generally true when the state-changing transformations are elements of the Virasoro algebra, however it holds for the transformations that we consider in the main text. The results of Section~\ref{sec:transport} utilize the formula for the curvature presented in this appendix.

Consider a Lie algebra ${\mathfrak g}$ and a trajectory of elements $X(\lambda)\in {\mathfrak g}$ specified by some parameter $\lambda$. We write ${\rm Ad}_X$ for the adjoint action of $X$ on ${\mathfrak g}$, ${\rm Ad}_X(Y)=[X,Y]$. We make the assumption that the kernel of ${\rm Ad}_X$ and the image of ${\rm Ad}_X$ do not intersect anywhere along the path, which is guaranteed if $[X,Y]\neq 0$ implies $[X,[X,Y]]\neq 0$. Moreover, we will be interested in smooth trajectories $X(\lambda)$ along which the kernel and image of ${\rm Ad}_X$ vary smoothly. In particular, we will assume their dimensions do not jump. 

Crucially, we will make the further assumption\footnote{For finite-dimensional Lie algebras the dimensions of the kernel and the image add up to the total dimension of the Lie algebra. Since they do not intersect, this then implies that the kernel and image of ${\rm Ad}_X$ together span the full Lie algebra. For infinite-dimensional Lie algebras the situation is more complicated, as we explain in Appendix \ref{sec:NonDiagonalization}.} that any $Y$ can be uniquely decomposed as
$Y=K+I$ with $K$ in the kernel and $I$ in the image of ${\rm Ad}_X$. We will call the corresponding projection operators $P_K$ and $P_I$, with the property that
\be \label{eq:projectrors}
P_I+P_K  =   1~.
\ee
Notice that we are not using an inner product, which means that the projectors are not orthogonal in any sense.

Besides the projectors $P_K$ and $P_I$, we will denote ${\rm Ad}_X$ simply by $A$, and its inverse by $A^{-1}$. Note that $A$ has a kernel so it does not have an inverse, but since by assumption $A$ defines a non-degenerate map from the image of the image of ${\rm Ad}_X$
to itself, it does have a well-defined inverse on these subspaces. The map $A^{-1}$ is defined to be the inverse on these subspaces and zero everywhere else. These operators then obey the following set of identities:
\bea\label{eq:identities}
A P_K = P_K A & = & 0~, \\ \label{eq:identity1}
A^{-1} P_K = P_K A^{-1} & = & 0~,  \label{eq:identity2}\\
AA^{-1} = A^{-1} A & = & P_I~. \label{eq:identity3}
\eea
We now vary $X$ to $X+\delta X$ by some small change $\delta \lambda$ along the path. In particular, we can use the above identities to express the variations of $P_K$, $P_I$ and $A^{-1}$ in terms of the variation of $A$. After some algebra we find that 
\bea \label{deltaPK}
\delta P_K &=& -\delta P_I  =  -P_K \delta A A^{-1} P_I - P_I A^{-1} \delta A P_K~,  \\
\delta A^{-1} & = & -A^{-1} \delta A A^{-1} + P_I A^{-2} \delta A P_K + P_K \delta A A^{-2} P_I~. \label{deltaAinverse}
\eea
In particular, we used 
\be
P_K\delta P_I = P_K\delta A^{-1}A P_I~, \quad P_I\delta P_I =P_I A^{-1}\delta A P_K~,
\ee
for deriving Eq.~\eqref{deltaPK} and
\be
P_K\delta A^{-1} =\delta P_I A^{-1}~, \quad P_I\delta A^{-1} = A^{-1}\delta P_I -A^{-1}\delta A A^{-1}~,
\ee
for Eq.~\eqref{deltaAinverse}. We also used that $P_I A^{-1}=A^{-1}P_I=A^{-1}$ and $P_I A=AP_I=A$.

Given a variation $\delta X$, we want to express it as
\be
\delta X = [S,X] + P_K\delta X~, \label{eq:decomposition}
\ee
where $P_K\delta X$ is in the kernel of ${\rm Ad}_X$. Moreover, we want to remove the modular zero mode from $S$, so that $S$ is uniquely defined.
We do this by requiring that $P_K S = S P_K=0$, and with the above equations it is then easy to see that
\be
S=-A^{-1}(\delta X)~.
\ee

We are now going to compute the parallel transport along a small square, by first doing the variation $\delta_1 X$ and then
$\delta_2 X$, and then subtracting the reverse order. For the difference, we get 
\be
F= (1-(A^{-1}+\delta_1 A^{-1})(\delta_2 X)) (1 - A^{-1}(\delta_1 X)) - (1 \leftrightarrow 2)~.
\ee
The first order terms vanish, thus it is necessary to expand to second order. One term we get
at second order is
\be
F_1 = - [A^{-1}  (\delta_1 X), A^{-1} (\delta_2 X)]~.
\ee
There is also another term coming from the variations of $A^{-1}$, which evaluates to
\be \label{eq:F2}
F_2 = (A^{-1} \delta_1 A A^{-1} -
P_I A^{-2} \delta_1 A P_K - P_K \delta_1 A A^{-2} P_I  
)(\delta_2 X) -  (1 \leftrightarrow 2)~.
\ee
In order to simplify Eq.~\eqref{eq:F2} further, we need several other identities. For example, multiplying
\be \label{eq:identity3-1}
A([Y,Z])=[AY,Z]+[Y,AZ]
\ee
by $A^{-1}$ we get the identity
\be \label{eq:identity3-2}
A^{-1}([AY,Z]+[Y,AZ])=P_I([Y,Z])~.
\ee
From this it follows that
\be \label{eq:identity4}
A^{-1} [Y,P_K Z] = A^{-1}[P_I Y, P_K Z] = P_I([A^{-1}Y,P_K Z])~,
\ee
where we used Eqs.~\eqref{eq:projectrors}, \eqref{eq:identity1} and \eqref{eq:identity3}.

Next we consider the first term in $F_2$ minus the same term with $1$ and $2$ interchanged. It is given by
\be 
F_2^1 =A^{-1} \delta_1 A A^{-1}(\delta_2 X) - (1 \leftrightarrow 2)~.
\ee
We use $\delta_1 A Y=[\delta_1 X,Y]$ to rewrite it as 
\bea
F_2^1 & = & A^{-1} ([\delta_1 X,A^{-1}(\delta_2 X)] + [A^{-1}(\delta_1 X),\delta_2 X]) \nonumber \\
& = & A^{-1} ([(AA^{-1}+P_K)\delta_1 X,A^{-1}(\delta_2 X)] + [A^{-1}(\delta_1 X),(AA^{-1}+P_K)\delta_2 X]) \nonumber \\
& = & P_I ([A^{-1}(\delta_1 X),A^{-1}(\delta_2 X)]) + A^{-1}([P_K \delta_1 X,A^{-1}(\delta_2 X)] + 
 [A^{-1}(\delta_1 X),P_K\delta_2 X])~. \nonumber\\
\eea
In the last equality we make use of Eq.~\eqref{eq:identity3-2}. Applying Eq.~\eqref{eq:identity4} to the last two terms gives 
\be
F_2^1=P_I ([A^{-1}(\delta_1 X),A^{-1}(\delta_2 X)] + [A^{-2}(\delta_1 X),P_K\delta_2 X]- [A^{-2}(\delta_2 X),P_K \delta_1 X] )~.
\ee
The second term in $F_2$ reads
\bea
F_2^2 &=& - P_I A^{-2} \delta_1 A P_K(\delta_2 X) + P_I A^{-2} \delta_2 A P_K(\delta_1 X) \nonumber \\
&=&-
A^{-2}([\delta_1 X,P_K \delta_2 X] - [\delta_2 X,P_K \delta_1 X])~.
\eea
Using the identity Eq.~\eqref{eq:identity4} twice it follows that
\be
F_2^2 = - P_I
([A^{-2}(\delta_1 X),P_K \delta_2 X] - [A^{-2} (\delta_2 X),P_K \delta_1 X])~.
\ee
The last term to consider is
\bea
F_2^3 &=&P_K \delta_2 A A^{-2} P_I  
(\delta_1 X) - P_K \delta_1 A A^{-2} P_I  
(\delta_2 X) \nonumber \\
&=& P_K([\delta_2 X, A^{-2}(\delta_1 X)]-[\delta_1 X, A^{-2}(\delta_2 X)] )~.
\eea
This expression does not admit an obvious simplification. Combining all terms we see that the first term in $F_2^1$ cancels part of $F_1$, the second and third terms in $F_2^1$
cancel against $F_2^2$, so that we are left with a simple and compact expression for the full curvature:
\be \label{eq:curv}
F = - P_K(  [A^{-1}  (\delta_1 X), A^{-1} (\delta_2 X)] + [\delta_1 X, A^{-2}(\delta_2 X)] - [\delta_2 X, A^{-2}(\delta_1 X)])~.
\ee
One can easily check that the curvature commutes with $X$. 

Notice that only the $P_I$ components of $\delta X$ contribute to the curvature due to the 
observation that
\be \label{eq:PKPIPK}
P_K([P_I Y, P_K Z]) =  P_K([A A^{-1} Y, P_K Z])=P_K A([A^{-1}Y,P_K Z])=0~,
\ee
where we used Eq.~\eqref{eq:identity3-1}. Moreover, we find that 
\bea
W &=& A^2 ([A^{-2} (\delta_1 X), A^{-2} (\delta_2 X)]) \nonumber \\
&=& 2 [A^{-1}  (\delta_1 X), A^{-1} (\delta_2 X)] + 
[P_I \delta_1 X, A^{-2}(\delta_2 X)] + [ A^{-2}(\delta_1 X),P_I \delta_2 X]
\eea
is almost the same as Eq.~\eqref{eq:curv}, except for the factor of two, and the appearance of the projector $P_I$. It is obvious that $P_KW=0$ and if we add $P_KW$ to $F$ we can drop the $P_I$ in the resulting expression, as follows from Eq.~\eqref{eq:PKPIPK}. Therefore, the final expression for the curvature reads
\be
F = P_K(  [A^{-1}  (\delta_1 X), A^{-1} (\delta_2 X)] )~.
\ee
The simple form of this result
suggests that there is a shorter derivation and it would be interesting to further investigate this possibility.

\section{Non-diagonalization for Virasoro}\label{sec:NonDiagonalization}

There are subtleties in expressing a Virasoro generator $X$ as $X=X_0+[H_{\rm mod},Y]$ with $X_0$ a zero mode of the modular Hamiltonian $H_{\rm mod}$ in the Virasoro algebra. We will give here a summary of why the assumed decomposition, Eq.~\eqref{eq:decomposition}, used to derive the curvature cannot be applied to the full Virasoro algebra, and hence why we have chosen to restrict to a different set of transformations.

We will first be more precise about the notion of `generator.' A generator of Diff($S^1$) can be expressed as 
\be \label{eq:generator}
X=\sum_n c_n L_n~,
\ee
where the modes $L_n$ satisfy the Virasoro algebra, Eq.~\eqref{eq:virasoroalgebra}.
We can equivalently represent $X$ as a function on $S^1$, $f(\theta)=\sum c_n e^{in\theta}$, or as a vector field,
$\xi = \sum c_n z^{n+1} \partial_z$ in radial quantization. For the arguments we are interested in the central charge can be considered separately, see Section~\ref{sec:extension}.

One can ask what values of $c_n$ are allowed in Eq.~\eqref{eq:generator}. This leads to different `definitions' of the Virasoro algebra. Some choices that are preserved under commutation are:
\begin{itemize}
\item algebraic: require only a finite number of the $c_n$ to be non-zero~,
\item semi-algebraic: require that $c_n=0$ for $n$ sufficiently negative (alternatively, one could require $c_n=0$ for $n$ sufficiently positive)~,
\item analytic: require the function $f$ or vector field $\xi$ to be smooth~.
\end{itemize}
In the case where the generators are self-adjoint, then semi-algebraic reduces to algebraic. 

For each of these choices of infinite-dimensional Lie algebras, we can ask to what extent the statement that any generator $X$ can be written 
as $X=X_0+[H_{\rm mod},Y]$ with $X_0$ a zero mode of the modular Hamiltonian $H_{\rm mod}$ holds.

\subsection{Algebraic and semi-algebraic case}

In the algebraic case, one can prove that the only algebra element that commutes with $H_{\rm mod}$ is $H_{\rm mod}$ itself. First, recall that 
\be H_{\rm mod}=\pi(L_1+L_{-1})~.\ee 
Now consider elements with only a finite number of non-zero $c_n$, running from $n=-L,...,K$, with $K$ and $L$ positive. Then, the commutator  
\be
[H_{\rm mod},\sum_{n=-L}^{K} c_n L_n] = \sum_{n=-L-1}^{K+1} c'_n L_n~
\ee
maps a vector space of dimension $K+L+1$ into a vector space of dimension $K+L+3$. Its kernel is dimension one so its cokernel 
must be dimension three. Therefore, the number of generators which can be written as $[H_{\rm mod},X]$ is codimension three. In fact, one can write every generator as
\be  \label{j1}
X = aH_{\rm mod} + bL_2 + cL_{-2} + [H_{\rm mod},Y]~,
\ee
for some $a,b,c$, which can be seen iteratively by taking a suitable $Y$ with $L=K=1$ and combining $H_{\rm mod}, L_2, L_{-2}$ to isolate $L_0$, then taking a suitable $Y$ with $L=1,K=2$ combined with all the previous generators to isolate $L_3$, and so on and so forth. Crucially, this decomposition is not unique. For instance, we could have equally well written a similar decomposition with $L_3,L_{-3}$ instead of $L_2,L_{-2}$.

To solve
\be 
L_{-2} = [H_{\rm mod},Y]~,
\ee
it is necessary to express $Y$ as an infinite series $Y=\sum_{k=-3}^{-\infty} c_k L_k$ which is not part of the algebra:
\be
Y= \frac{1}{4} L_{-3} - \frac{2}{4\cdot 6} L_{-5} + \frac{2}{6 \cdot 8} L_{-7} - \frac{2}{8 \cdot 10}L_{-9} + \ldots
\ee
If we denote by $Y_k$ the sum of the first $k$ terms which truncates at $L_{-2k-1}$, then we have
\be
\frac{1}{\pi}[H_{\rm mod},Y_k]=L_{-2} +\frac{(-1)^{k+1}}{k+1} L_{-2k-2}~,
\ee
so that for large $k$ this becomes `close' to $L_{-2}$. We can introduce a metric so that this notion of closeness becomes more precise, e.g.,
\be
|| \sum_n c_n L_n ||^2 \equiv \sum_n |c_n|^2\label{eq:norm}
\ee
defines a metric on the Lie algebra. But the Lie algebra is not complete with respect to this metric, i.e., limits of Lie algebra elements
which converge in this norm will not in general converge to an element of the Lie algebra. 

Even ignoring the fact that the algebra is not complete with respect to Eq.~\eqref{eq:norm}, there is the additional issue that this way of interpreting $L_{-2}$ as the commutator of an element of the algebra with $Y$ is too strong. Indeed, we can also find an infinite series $Y$ obeying
\be
[H_{\rm mod},Y]=H_{\rm mod}~,
\ee
which looks like 
\be
Y=\ldots + c_6 L_6 + c_4 L_4 + c_2 L_2 + c_{-2} L_{-2} + c_{_4} L_{-4} + c_{-6} L_{-6} +\ldots
\ee
This also has the property that if one truncates $Y$, the $Y_k$ obeys $[H_{\rm mod},Y_k]=H_{\rm mod}+Z_k$, with $Z_k$ small defined with respect to the above norm. This would not allow for a decomposition separating out the zero mode part from the image of the adjoint action without intersection.

Notice that considering the semi-algebraic rather than algebraic case also does not fix the issue. A semi-infinite series in one direction can either remove $L_2$ or $L_{-2}$ 
from the expression Eq.~(\ref{j1}), but not both.

\subsection{Analytic case}
In the analytic case, the equation $[H_{\rm mod},X]=Y$ is the differential equation
\be\label{eq:analyticdiffeq}
(1+z^2)X'(z) -2z X(z) = -\frac{1}{\pi}Y(z)~,
\ee
where we replaced everything by the corresponding smooth function. This differential equation is equivalent to
\be
\frac{d}{dz} \left( \frac{X(z)}{1+z^2} \right) = -\frac{1}{\pi}\frac{Y(z)}{(1+z^2)^2}~.
\ee
Therefore,
\be \label{eq:X(z)}
X(z) = -\frac{c_0}{2}(1+z^2) -\frac{1}{\pi}(1+z^2) \int^z \frac{Y(z')}{(1+z'^2)^2}\, dz'~,
\ee
where $c_0$ is an integration constant, and the integration is over the circle. The differential equation does not have an analytic solution for all $Y(z)$. In fact, we will argue that in order to find an analytic solution we require three conditions on $Y$, so that once again the space of smooth vector fields which can be written as $[H_{\rm mod},X]$ is codimension three. 

The first two conditions come from exploring the behavior of the integrand near $z=\pm i$, where we find that there will be logarithmic branch cut singularities unless the residues at $z=\pm i$ vanish. Thus, the first two conditions on $Y(z)$ for Eq.~\eqref{eq:X(z)} to be analytic are
\be\label{eq:vanishresidue}
{\rm Res}_{z=\pm i} \, \frac{Y(z)}{(1+z^2)^2} =0~.
\ee
Note that it is admissible for $Y(z)/(1+z^2)^2$ to have double pole at $z=\pm i$, as these integrate
to a single pole, which is then canceled by the $(1+z^2)$ prefactor in Eq.~\eqref{eq:X(z)}. Therefore, the double poles do not give rise to singularities. 

There is also another condition, namely that the contour integral of $X'(z)$ around the unit circle vanishes so that we get a periodic
function $X(z)$ after integration. Since polynomials in $z$ are automatically periodic, it suffices to consider the behavior of the integrand, $Y(z)/(1+z^2)^2$. Assuming that $Y(z)$ is analytic except possibly at $z=0$, this amounts to the condition
\be \label{eq:third condition}
{\rm Res}_{z=0} \, \frac{Y(z)}{(1+z^2)^2} =0~.
\ee
Note that poles near $z=\pm i$ do not affect the periodicity so we can subtract them before applying this condition if necessary, and we also assume the residues vanish as above, so that we have a well-defined integral.

To see how this works in practice, it is useful to evaluate this for a trial function $Y$ inspired by the algebraic case:
\be
Y_0 = a(1+z^2) + b z^{-1} + c z^3~,
\ee
which contains $L_2$, $L_{-2}$ and $H_{\rm mod}$. We notice that
\begin{align}
\frac{Y_0}{(1+z^2)^2} &= \frac{i(b+c)}{4(z-i)^2} +\frac{-b-ia+c}{2(z-i)} + \ldots~, \\
\frac{Y_0}{(1+z^2)^2} &= \frac{-i(b+c)}{4(z+i)^2} +\frac{-b+ia+c}{2(z+i)} + \ldots
\end{align}
near $z=\pm i$ respectively. The residue of $Y_0/(1+z^2)^2$ at $z=i$ equals $-(iY_0(i)+Y_0'(i))/4$ and the residue at $z=-i$ equals 
$(iY_0(-i)-Y_0'(-i))/4$, and these are required to vanish by Eq.~\eqref{eq:vanishresidue}. This translates to $b=c$ and $a=0$. Recall that the differential equation, Eq.~\eqref{eq:analyticdiffeq}, extracts the non-zero mode part, i.e., the vector fields that can be written as $[H_{\rm mod}, X]$. We could also ask how to extract the zero mode part. In this case it seems the most natural choice to extract $a$, which is given by the difference of the two residues, as the coefficient of the zero mode. 

Even in the case $b=c$ and $a=0$ with vanishing residues, we see that $X$ will now involve a term $(1+z^2)\log z$ since $Y=z^{-1}+z^3=(z^2+1)^2z^{-1}-2z$. This still has a branch cut singularity, and therefore will not be single-valued. This is where a version of the third condition, Eq.~\eqref{eq:third condition}, is necessary. To be more precise about this requirement, take a finite polynomial in $z,z^{-1}$ for $Y$.
We first subtract the harmless double poles and the harmful single poles (which we require to vanish independently) so that we get
an expression of the type
\be \label{j3}
Z(z)\equiv\frac{Y(z) - A - Bz -Cz^2- Dz^3}{(1+z^2)^2}~,
\ee
where the coefficients $A,B,C,D$ are chosen so as to cancel the single and double poles. To accomplish this, it is necessary for an overall factor $(1+z^2)^2$ to factor out of the numerator. The choice of coefficients can then be determined by the requirement that the numerator of $Z$ and its derivative both vanish at $z=\pm i$.
Explicitly, they are given by
\bea \label{j4}
A & = & \frac{1}{4}(2Y(-i)+2Y(i)+iY'(-i)-iY'(i))~, \\
B & = & \frac{1}{4}(3iY(-i)-3iY(i)-Y'(-i)-Y'(i))~, \\
C & = & \frac{i}{4}(Y'(-i)-Y'(i))~,  \\
D & = & \frac{1}{4}(iY(-i)-iY(i)-Y'(-i)-Y'(i))~. \label{eq:D}
\eea
With this choice of coefficients the expression, Eq.~\eqref{j3}, is now
well-behaved everywhere, i.e., the numerator has a factor $(1+z^2)^2$, and the quotient is also a finite polynomial in $z$ 
and $z^{-1}$. The only problematic contribution to the integral is coming from the $z^{-1}$ term which does not become a periodic function
when integrated. So the remaining number is the coefficient in front of $z^{-1}$ in the polynomial $Z(z)$ in Eq.~(\ref{j3}).

We denote by $Y_-$ the terms in $Y$ with a negative power of $z$. The non-negative powers in $Y$ only give rise to non-negative powers
in $Z$ and are never problematic. So we can equivalently consider 
\be \label{j5}
Z_-(z)\equiv\frac{Y_-(z) - A - Bz -Cz^2- D z^3}{(1+z^2)^2}~,
\ee
and we are interested in the coefficient in front of $z^{-1}$ in $Z_-(z)$. We can extract this using a small contour integral.
But we might as well extract it using a large contour integral as $Z_-$ is analytic everywhere except at $0$ and $\infty$. Then the integral is dominated by $D$, so it is necessary that $D=0$ for the integral to be single-valued. In fact, $D$ is equal to the sum of the residues at $z=i$ and $z=-i$, as can be seen from Eq.~\eqref{eq:D}, so this version of the third condition with the double poles subtracted out reduces to
\be
{\rm Res}_{z=i} \frac{Y_-}{(1+z^2)^2} + {\rm Res}_{z=-i} \frac{Y_-}{(1+z^2)^2} =0~.
\ee
Since the residues of the complete $Y/(1+z^2)^2$ have to vanish separately, we could equivalently require the same
condition for $Y_+$. 

For more general non-polynomial $Y$, we can apply the same argument, except that now $Y_-$ is analytic outside the unit disk and
$Y_+$ is analytic inside the unit disk. By the version of the Riemann-Hilbert problem that applies to simple closed curves, a decomposition of analytic functions on the circle of the type $Y_- + Y_+$ exists.

\subsection{Issues from non-diagonalization} \label{app:nondiag}
In this subsection, we will show that the ambiguities in the diagonalization of the Virasoro algebra with respect to the adjoint action translate to ambiguities in the projection operator. This leads to different answers for the Berry curvature that are physically inequivalent. As a result, there is no sensible bulk interpretation. It is because parallel transport acting by elements of the usual Virasoro algebra is plagued with ambiguities that we are forced to extend to a non-standard algebra constructed from vector fields on the half-circle as in Section~\ref{sec:transport}, where the construction is unique.

For the ordinary Virasoro case, we want to construct a zero-mode projector $P_0$ so that it evaluates to zero for the integrand of Eq.~\eqref{eq:nonzeromode1}, while it gives a non-zero value for Eq.~\eqref{eq:Hmodplane}. 
In other words we can devise a contour integral prescription in such a way as to satisfy the properties:
\begin{itemize}
    \item The functional is non-zero on the modular Hamiltonian, i.e., $P_0\left(H_{\rm mod}\right)=1$~,
    \item It projects out the commutator of the modular Hamiltonian with anything else, i.e., $P_0\left([H_{\rm mod},X_{\xi}]\right)=0~$, for any vector field $\xi(z)$~.
\end{itemize}
We emphasize that this is a different projection operator than the one considered in Section~\ref{sec:transport}, in particular it is finite rather than a delta function.

There are several different choices that obey both of these properties:
\begin{align}
P^{(1)}_0(X_{\xi})&\equiv-\frac{1}{\pi^2}\int_{|z+i\epsilon|=1}\frac{\xi(z)}{(1+z^2)^2}\,dz~,\\
P^{(2)}_0(X_{\xi})&\equiv\frac{1}{\pi^2}\int_{|z-i\epsilon|=1}\frac{\xi(z)}{(1+z^2)^2}\,dz~,\\
P^{(3)}_0(X_{\xi})&\equiv \frac{1}{2}\left(P^{(1)}_0(X_{\xi})+P^{(2)}_0(X_{\xi})\right)~. \label{eq:projop}
\end{align}
By explicitly computing the residues, one can express these in terms of the diffeomorphism $\xi$ and its derivative evaluated at the endpoints of the interval as
\begin{align} \label{eq:residue}
P^{(1)}_0(X_{\xi})&=\frac{1}{2\pi}\left[\xi(-i)+i\xi'(-i)\right]~,\\
P^{(2)}_0(X_{\xi})&=\frac{1}{2\pi}\left[\xi(i)-i\xi'(i)\right]~,\\
P^{(3)}_0(X_{\xi})&=\frac{1}{4\pi}\left[i\,\xi'(-i)-i\,\xi'(i)+\xi(-i)+\xi(i)\right]~.\label{eq:projop2}
\end{align}
Note that the sum of contours $P_0^{(2)}-P_0^{(1)}$ does not satisfy the required properties, as it vanishes on the modular Hamiltonian. The difference of contours, Eqs.~\eqref{eq:projop} and \eqref{eq:projop2}, is perhaps the most symmetrical choice. It can be seen to result from the decomposition, Eq.~\eqref{j1}, by additionally imposing that the linear functional evaluated on the extra terms $L_2,L_{-2}$ in the decomposition give zero. However, recall that this decomposition was not unique. A different choice would have resulted in a different linear functional, and therefore a different $P_0$.

Moreover, we have considered the possibility of defining a zero mode projector $P_0$ using very early or very late time modular flow. However, we found that this prescription is also ambiguous and depends on whether one considers very early or very late times.

\begin{figure}[t!]
\centerline{\includegraphics[scale=.5]{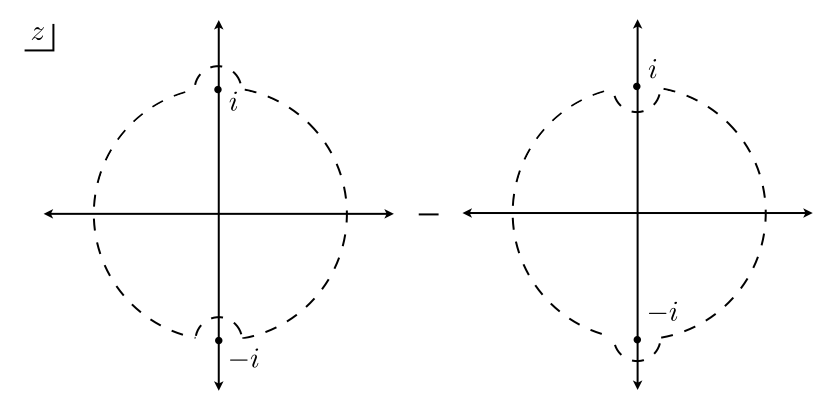}}
\caption{One simple choice of linear functional, constructed from the difference of $|z-i\epsilon|=1$ and $|z+i\epsilon|=1$ contours. When considering a non-restricted set of generators, there is an ambiguity in the choice of projection. For instance, it is also possible to choose either of these contours separately (but not their sum) and still satisfy the required properties for the linear functional. This ambiguity is tied to the fact that the adjoint action is not diagonalizable over the Virasoro algebra.}
\label{fig:contour}
\end{figure}

It is also easy to see that this has a direct physical implication by leading to different results for the curvature. For instance, consider an infinitesimal diffeomorphism of the form 
\begin{equation}\label{conformal transformation}
     \theta\rightarrow \theta +2\epsilon \sin{(m \theta)}~,
\end{equation}
where $m \in \mathbb{Z}$. The parameter $\epsilon$ is assumed to be small and dimensionless. 

One can consider a parallel transport process consisting of a series of such infinitesimal transformations, where $m$ can vary from step to step. It is described by a function $m(\lambda)$, where $\lambda$ denotes the point along the path evaluated in the continuum limit. 

Mapping from the cylinder to the plane using Eq.~\eqref{conformal transformation} and expanding to first order in $\epsilon$, this sinusoidal perturbation becomes
\begin{align}
 \xi(z) =z+\epsilon(z^{m+1}-z^{-m+1})+\mathcal{O}(\epsilon^2)~. \label{conformal transformation 2}
\end{align}
Up to terms that are higher order in $\epsilon$, Eq.~\eqref{conformal transformation 2} can be inverted to $z = \xi - \epsilon(\xi^{m+1}-\xi^{-m+1}) + \mathcal{O}(\epsilon^2)$. Inserting this in Eq.~\eqref{eq:Hmodplane} for $H_{\rm mod}$, we find the correction to the modular Hamiltonian:
\begin{equation}
    H^{(1)}=\pi\left[(m+1)(L_{-m+1}+ L_{m-1}) +(m-1)(L_{-m-1}+L_{m+1})\right]~.
\end{equation}

Recall that expanding both the parallel transport equation $H_{\rm mod}=[S,H_{\rm mod}]$ order by order in $\epsilon$ gave Eq.~\eqref{eq:transportfirst}.
Solving for the correction to the parallel transport operator gives
\begin{equation}
    S^{(1)}= L_m-L_{-m}~, \indent S^{(0)}=0~.
\end{equation}
Take two transformations of the form Eq. \eqref{conformal transformation 2}  with different values for the integer $m$, say $m_1$ and $m_2$. This gives two different parallel transport operators, $S_1$ and $S_2$. To compute the curvature, Eq.~\eqref{eq:berrycurvature}, we are interested in computing the commutator 
\be \label{eq:commutator}
[S^{(1)}_1-\kappa_1H^{(0)},S^{(1)}_2-\kappa_2H^{(0)}]~,
\ee
where $\kappa_i=P_0(S_i)$, is the zero mode coefficient of the parallel transport operator $S_i$. We can split Eq.~\eqref{eq:commutator} into terms that we can treat separately. Notice that the term proportional to $[H^{(0)},H^{(0)}]$ is zero and can be neglected. By definition, the projection operator vanishes on $[S^{(1)}_i,H^{(0)}]$, so this contribution to the curvature is zero. An explicit computation shows that
\be
    [S^{(1)}_1,S^{(1)}_2]= (m_1-m_2)(L_{m_1+m_2}-L_{-m_1-m_2})+(m_1+m_2)(L_{-m_1+m_2}-L_{m_1-m_2})~.
\ee

We will now project onto the zero modes of each of the terms. This is where the ambiguity enters since the result depends on the choice of linear functional. We find
\begin{align}
    F^{(1)} = P^{(1)}_0([S^{(1)}_1,S^{(1)}_2])&= \frac{2i}{\pi}(m_2^2-m_1^2)\sin\left(\frac{m_1\pi}{2}\right)\sin\left(\frac{m_2\pi}{2}\right)~,\\
    F^{(2)} = P^{(2)}_0([S^{(1)}_1,S^{(1)}_2])&= -F^{(1)}~,\\
    F^{(3)} = P^{(3)}_0([S^{(1)}_1,S^{(1)}_2])&= 0~.
\end{align}
Notice that in the case where the $m_1,m_2$ are even, all curvatures agree and in fact identically vanish. Indeed, it is possible to argue that the curvature defined in this way always vanishes for diffeomorphisms that vanish on the interval endpoint. However, in general they do not agree and the result is ambiguous.